# PRODUCTION AND NETWORK FORMATION GAMES WITH CONTENT HETEROGENEITY

Yu Zhang, Jaeok Park, and Mihaela van der Schaar

***Abstract***— *Online social networks (e.g. Facebook, Twitter, Youtube) provide a popular, cost-effective and scalable framework for sharing user-generated contents. This paper addresses the intrinsic incentive problems residing in social networks using a game-theoretic model where individual users selfishly trade off the costs of forming links (i.e. whom they interact with) and producing contents personally against the potential rewards from doing so. Departing from the assumption that contents produced by difference users is perfectly substitutable, we explicitly consider heterogeneity in user-generated contents and study how it influences users' behavior and the structure of social networks. Given content heterogeneity, we rigorously prove that when the population of a social network is sufficiently large, every (strict) non-cooperative equilibrium should consist of either a symmetric network topology where each user produces the same amount of content and has the same degree, or a two-level hierarchical topology with all users belonging to either of the two types: influencers who produce large amounts of contents and subscribers who produce small amounts of contents and get most of their contents from influencers. Meanwhile, the law of the few disappears in such networks. Moreover, we prove that the social optimum is always achieved by networks with symmetric topologies, where the sum of users' utilities is maximized. To provide users with incentives for producing and mutually sharing the socially optimal amount of contents, a pricing scheme is proposed, with which we show that the social optimum can be achieved as a non-cooperative equilibrium with the pricing of content acquisition and link formation.*

## I. INTRODUCTION

Online social networks (OSNs) provide a popular, cost-effective and scalable framework for sharing user-generated contents (e.g. knowledge, photos, videos, news). Well known functions and examples of OSNs include social interactions on Facebook and MySpace, file sharing on YouTube and Flickr, messaging and exchanging on Twitter, and professional networking on LinkedIn. Due to their popularity, OSNs have the potential to fundamentally change our social lives [1][2][3].

Creating and dissolving links to share knowledge and contents is a key feature of OSN services. A user's ability to visualize and traverse his own connections and the connections of others (such as friends and friends of friends) is a defining feature of many OSNs [5]. Typical examples of link creation include "send friend request" on Facebook, "add to circle" on Google+, "follow" on Twitter, and "subscribe" on YouTube. Tremendous efforts have been dedicated in recent years to analyzing and describing the emerging social interactions among users as well as to understanding why certain connectivity pattern emerges in the network [3]-[7]. For instance, [3] reveals that the empirical probability of users who are followed by very large numbers of users on Twitter is above what a scale-free distribution would predict, while the relation of "following" is reciprocated in only around 22% of the cases. Meanwhile, [4] shows that ties on Twitter also exhibit high directed closure: a user will be followed by the followers of his followers with a high probability. The work of [6] records network evolution on Flickr and Yahoo! 360. It shows that a large fraction of the individuals are either isolated singletons or form part of small components. Similarly, empirical studies have been performed in [7] on the formation of friendship links on a music sharing site, which suggests that the formation of friendship ties is consistent with users' rational linking choices.

These empirical findings have also attracted the attention of theoretical microeconomics researchers [8], who focus on how the users' self-interests leads to strategic link creation among the users in an network and analyze the network topologies that arise out of purposeful individual actions. A simple model is analyzed in [9], where the problem of network formation on OSNs is formulated as a non-cooperative game among $n$ strategic users. The link creation actions are available to each user who aims to individually maximize his own utility by trading off the potential rewards obtained by forming a link (e.g. acquired contents) against the incurred link creation cost (e.g. payment and maintenance cost). The link formation is unilateral: a user $i$ can decide to connect with any user $j$ by paying for the link. Analysis of this model predicts the emergence of resulting "equilibrium" topologies, such as circles, stars, and variants of the star. This model also has been extended in a number of directions. For example, [10] studies the network formation problem when users are heterogeneous, where the costs and benefits of a link depends on the users it connects. It shows that a strict equilibrium network is minimal and conversely every minimal network is a strict equilibrium for suitable costs and benefits. [11] and [12] consider indirect content transmission, where users can access not only contents from friends (i.e. users directly connected via links), but also contents from friends of friends. They prove that if the benefit brought by contents is decaying in distance from the content source to the destination, users have incentive to get close to others. Meanwhile, the incremental value of contents falls as users acquire more contents. These findings result in small world networks [13] supported by a few links created by a great many people but pointing to only a few key users (the "influencers").

As in [9], our work also employs a non-cooperative one-shot game formulation to analyze the network formation in OSNs. However, different from previous works which assume that users are endowed with exogenous amounts of contents and only focus on the strategic aspect of link creation [9][10], our model explicitly considers the incentives of users to produce contents personally. The strategic connection between content production and link creation is also studied in the idealized work in [14], where the authors find that a phenomenon called "the law of the few" emerges as the result of strategic interactions among users. That is, in every strict equilibrium of the game, the network has a core-periphery architecture: the users in the core produce contents personally while the peripheral users produce no content personally but form links and get all their contents from the core users. Moreover, the population of the core users is upper-bounded which increase at a slower pace than the growth of the social group. Nevertheless, [14] assumes that for each user, contents produced by different users is equally valued and perfectly substitutable, which fails to capture content heterogeneity and users' desire for content diversity that exist in OSNs [15]. That is, a user's benefit from content consumption does not only depends on the total amount of contents he consumes, but also on how many different types of content and what amount of each type he acquires. We explicitly consider such heterogeneity in user-generated contents in our analysis and design by using the model of public goods introduced in [16].

In contrast to [14], we show under our model with content heterogeneity, an OSN possesses the following properties:

(1) **Equilibrium.** When the size (population) of an OSN is sufficiently large, every (strict) non-cooperative equilibrium should consist of either a symmetric topology where each user produces the same amount and has the same degree, or a hierarchical topology with all users belonging to either of the two types: *influencers* who produce large amounts of contents and *subscribers* who produce small amounts of contents and get most of their contents from influencers. Nevertheless, under content heterogeneity, "the law of the few" disappears in OSNs as the number of influencers grows proportionally with the network size and its fraction in the user population does not diminish to 0. Particularly, we also prove that a star topology can never emerge in any strict equilibrium. Therefore, production is no longer concentrated in a few powerful users but becomes more dispersed in OSNs.

(2) **Social Optimum.** We prove that the social optimum in an OSN is not necessarily achieved with a star topology as in [14], but can be achieved in a symmetric topology. To eliminate the efficiency loss between non-cooperative equilibria and the social optimum, we design a pricing scheme by charging for content acquisition and link creation, which align users' incentives to the maximization of social welfare and the social optimum can be achieved at a non-cooperative equilibrium.

The remainder of this paper is organized as follows. In Section II, we describe our basic model of content production and link creation, where the content heterogeneity is explicitly formulated. In Section III, we analyze the non-cooperative equilibria of this model. In Section IV, we analyze the social optimum of this model and propose a pricing scheme to achieve the social optimum at a non-cooperative equilibrium. We conclude in Section V with future research also outlined.

## II. SYSTEM MODEL

We consider the content sharing in a network where individual users choose to personally produce contents and to form connections with others to acquire the contents which they produce. Examples of such contents are news, photos, expert knowledge, restaurant reviews, job information, political opinions, and etc. We construct an abstract model of content production and link creation in such networks, which we refer to as the production and network formation game (PNF game).

Let $N = \{1, 2, \dots, n\}$ denote the set of users in the network where $n \geq 3$ and let $i$ and $j$ denote typical members in this set. Each user determines his level of production. $x_i \in \mathbb{R}^+$ represents the amount of contents produced by user $i$. We assume that a user has no private contents preserved and hence, $x_i$ is also the amount of contents that user $i$ makes available for other users to acquire. User $i$ determines the set of users with whom he creates links, i.e. he "*subscribes*" to the contents of those users, which is represented by a row vector $\boldsymbol{g}_i = \left(g_{i1}, \dots g_{ii-1}, g_{ii+1}, \dots, g_{in}\right) \in \{0,1\}^n$ [1]. He subscribes to user $j$, if and only if $g_{ij} = 1$. For convenience, we set $g_{ii} = 0$ for all $i \in N$. The friend status between two users $i$ and $j$ is represented by $\overline{g}_{ij} = \max\{g_{ij}, g_{ji}\}$, which indicates that these two users are being friends if either of them subscribes to each other. We assume that the content sharing between two friends is bilateral. Hence, each user can acquire and consume all contents produced by his friends. This models the content sharing feature on popular OSNs such as Facebook and LinkedIn [2].

The set of strategies of user $i$ is thus denoted by $S_i = \mathbb{R}^+ \times \{0,1\}^n$. Let $\boldsymbol{x} = \left(x_i\right)_{i=1}^n$ and $\boldsymbol{g} = \left(\boldsymbol{g}_i\right)_{i=1}^n$ denote the production and subscription graphs of an OSN, a strategy profile is $\boldsymbol{s} = \left(\boldsymbol{x}, \boldsymbol{g}\right) \in S = \prod_{i \in N} S_i$. Similarly, we define $\overline{\boldsymbol{g}} \triangleq \left[\overline{g}_{ij}\right]_{i,j \in N}$ as the friend graph of the network. Let $N_i\left(\boldsymbol{g}\right) \triangleq \{j \mid g_{ij} = 1\}$ denote the set of user $i$'s subscriptions and $N_i\left(\overline{\boldsymbol{g}}\right) \triangleq \{j \mid \overline{g}_{ij} = 1\}$ denote the set of user $i$'s friends, his degree is thus represented as $d_i\left(\overline{\boldsymbol{g}}\right) \triangleq \left|N_i\left(\overline{\boldsymbol{g}}\right)\right|$.

Users in the PNF game benefit from consuming the contents that he can acquire (both from self-production and subscriptions). We assume pure local externalities [9][14]: each user only acquires the contents personally produced by himself and his friends [3]. In other words, a user cannot acquire the contents produced by another user who is more than one hop away from him on the friend graph $\overline{\boldsymbol{g}}$.

---

[1] There are also alternative models about a user's linking strategy. For instance, the linking action $g_{ij}, \forall i, j$ could be a real number within the range $[0,1]$ instead of being binary, which represents the strength of the link that user $i$ forms with user $j$.

[2] However, there are also some social networking services (e.g. Twitter) where the content sharing is unilateral. For instance, a user can only view the contents shared by other users whom he subscribes to (i.e. the "follow" activity). The analysis in this work can also be applied into this scenario.

[3] An analysis about indirect content transmission can be found in [11].

Most existing works on network formation games assume that for each user, the contents produced by different users are perfectly substitutable in consumption [9][14], i.e. the total amount of contents that a user consumes fully determines his benefit. In other words, a user's benefit from content consumption will remain unchanged if he reduces his production by a certain amount and acquires the same amount of extra contents from another user, and vice versa. This assumption, however, fails to capture the content heterogeneity and users' interests to consume diverse contents. Hence, instead of assuming perfect substitutability, we use the preference model from [16] to capture the heterogeneity among different classes of contents. Particularly, we assume that each user has certain expertise in producing a particular class of contents and a user $i$'s benefit from content consumption is given by

$$f\left(x_i, \boldsymbol{x}_{-i}, \boldsymbol{g}\right) = v\left(\left[x_i^{\rho} + \sum_{j \in N_i(\bar{g})} x_j^{\rho}\right]^{\frac{1}{\rho}}\right). \tag{1}$$

where we use the convention that $x_i$ and $\boldsymbol{x}_{-i}$ refer to the production levels of user $i$ and of all users other than $i$. $\rho \in (0,1)$ measures a user's desire for content diversity. With $\rho < 1$, it is not the total amount of contents consumed but the number of different types of contents acquired and the particular amount of each type that jointly determine a user's benefit from content consumption. A smaller value of $\rho$ indicates a higher level of users' desire for content diversity. When $\rho \to 1$, contents from different users becomes perfect substitutable, which degenerates our model to the model in [14]. For notational convenience, we define $X_i \triangleq \left[x_i^{\rho} + \sum_{j \in N_i(\bar{g})} x_j^{\rho}\right]^{\frac{1}{\rho}}$ as user $i$'s perceived amount of contents and

$$e\left(x_i, X_i\right) \triangleq \frac{\partial f}{\partial x_i} = v'\left(X_i\right)\left(\frac{X_i}{x_i}\right)^{1-\rho}. \tag{2}$$

as his marginal benefit of production.

The following constraints are imposed on the benefit function (1):

*Assumption 1*: $v(\bullet)$ is a twice continuously differentiable, increasing, and strictly concave function.

*Assumption 2*: $v(\bullet)$ satisfies $v(0) = 0$, $v'(\bullet) > 0$ on $\mathbb{R}^+$, $v'(0) < \infty$, $v'(0) > \alpha$ where $\alpha > 0$ is a constant, and $\lim_{x \to \infty} v'(x) = 0$.

*Assumption 3*: $\left.\frac{\partial^2 f}{\partial x_i \partial x_j}\right|_{x_i, x_j \in \mathbb{R}^+} < 0,\ \forall i \ and \ \forall j \in N_i\left(\bar{\boldsymbol{g}}\right)$.

We briefly discuss these assumptions. The first two assumptions capture the saturation effect in a user's content consumption. That is, a user's benefit increases with the amount of contents he consumes, while the marginal benefit of production decreases and approaches to 0. Hence, there always exists an upper bound on the amount of information that an individual user can produce. Assumption 3 formalizes the externality among users' productions. User $i$'s incentive of production decreases against the amount of his acquisition from friends, as his marginal benefit of production decreases.

With the above three assumptions, $f\left(x_i, \boldsymbol{x}_{-i}, \boldsymbol{g}\right)$ is also twice continuously differentiable, increasing and strictly concave in $x_i$, as proved in the following lemma.

**Lemma 1.** $f\left(x_i, \boldsymbol{x}_{-i}, \boldsymbol{g}\right) = v\left(\left[x_i^\rho + \sum_{j \in N_i\left(\bar{\boldsymbol{g}}\right)} x_j^\rho\right]^{\frac{1}{\rho}}\right)$ *is twice continuously differentiable, increasing, and strictly concave in* $x_i$.

*Proof*: Taking the first-order derivative in $x_i$ over $f$, we have that

$$\frac{\partial f}{\partial x_i} = v'\left(X_i\right)\left(\frac{X_i}{x_i}\right)^{1-\rho}. \tag{3}$$

Since $v'\left(X_i\right) > 0$, we have that $\frac{\partial f}{\partial x_i} > 0$ and $f\left(x_i, \boldsymbol{x}_{-i}, \boldsymbol{g}\right)$ is strictly increasing in $x_i$.

Taking the second-order derivative, we have that

$$\frac{\partial^2 f}{\partial x_i^2} = v''\left(X_i\right)\left(\frac{X_i}{x_i}\right)^{2(1-\rho)} + v'\left(X_i\right)\left(1-\rho\right)\left(\frac{X_i}{x_i}\right)^{-\rho} \frac{\left(\frac{X_i}{x_i}\right)^{1-\rho} x_i - X_i}{x_i^2}. \tag{4}$$

Both the first and second terms in the RHS of (4) are smaller than 0. Hence, we have $\frac{\partial^2 f}{\partial x_i^2} < 0$ and thus $f\left(x_i\right)$ is twice differentiable and strictly concave in $x_i$. ■

We consider a linear cost on content production. That is, each user $i$ consumes a marginal cost of $c$ to produce one unit of contents. Hence, the cost of producing an amount $x_i$ is $cx_i$. Creating a link also incurs some fixed cost $\gamma$, which could be interpreted as the subscription fee that a user submits to the service provider of the network or the maintenance cost to keep a link [4]. The utility of user $i$ is given by his benefit of content consumption minus all costs:

$$u_i\left(\boldsymbol{x},\boldsymbol{g}\right) = v\left(\left[x_i^\rho + \sum_{j \in N_i\left(\bar{\boldsymbol{g}}\right)} x_j^\rho\right]^{\frac{1}{\rho}}\right) - cx_i - \gamma\left|N_i\left(\boldsymbol{g}\right)\right| \tag{5}$$

We analyze the case of homogeneous users in that $v$, $c$, $\gamma$, and $\rho$ are the same for all users. We assume that $\alpha > c$ to ensure the network is socially valuable. For illustration purposes, we also refer to $r_i\left(\boldsymbol{x},\boldsymbol{g}\right) \triangleq v\left(\left[x_i^\rho + \sum_{j \in N_i\left(\bar{\boldsymbol{g}}\right)} x_j^\rho\right]^{\frac{1}{\rho}}\right) - cx_i$ as user $i$'s *content utility*, which combines the user's benefit from content consumption and cost of content production.

## III. Equilibrium Analysis

### *A. Definition and basic properties*

In this section, we formalize the PNF game as a non-cooperative one-shot game. Each user maximizes his own utility given the strategies of others. A Nash equilibrium is defined as a strategy profile $\boldsymbol{s}^* = \left(\boldsymbol{x}^*, \boldsymbol{g}^*\right)$ such that

$$u_i\left(\boldsymbol{s}_i^*, \boldsymbol{s}_{-i}^*\right) > u_i\left(\boldsymbol{s}_i, \boldsymbol{s}_{-i}^*\right),\ \forall \boldsymbol{s}_i \in \mathbb{R}^+ \times \left\{0,1\right\}^n, \forall i \in N, \tag{6}$$

We explicitly consider strict equilibria and hence the inequality in (6) is set to be strict for every user. In the remainder of this paper, the words "equilibrium" and "strict equilibrium" are used interchangeably without further notice. We first analyze the basic properties of an equilibrium in the PNF game, and then apply the analysis separately to strategy profiles of symmetric production where each user produces the same amount of contents and strategy profiles of asymmetric production where users produce different amounts of contents.

[4] Our analysis could also be applied to the case when the subscription fee is paid to the user who is subscribed, where we have a different utility function **Error! Reference source not found.**

Given the equilibrium definition (6), a user $i$ 's equilibrium production $x_i^*$ always satisfies the following inequality:

$$e\left(x_i^*, X_i^*\right) = v'\left(X_i^*\right)\left(\frac{X_i^*}{x_i^*}\right)^{1-\rho} = c. \quad (7)$$

That is, the marginal benefit of production should equal to the marginal cost, and thus $i$ has no incentive to produce more than $x_i^*$ when his perceived amount of contents is $X_i^*$.

We first analyze the basic properties of users' behavior on content production and subscription when the network is in equilibrium. This is summarized in the following lemma.

**Lemma 2**. *In any equilibrium* $\left(\boldsymbol{x}^*, \boldsymbol{g}^*\right)$ *of the PNF game,*

*(1)* $g_{ij}^* g_{ji}^* = 0$ *for all* $i, j \in N$ *;*

*(2)* $x_i^* > 0$ *for all* $i \in N$ *.*

*Proof*: See Appendix A. ■

Lemma 2 is briefly explained as follows. In an equilibrium profile, each user's strategy is a strict best response to the strategies of others with his utility being maximized. Therefore, it leads to a redundant investment on link creation if there are two users who mutually subscribe to each other. Then, it can be concluded that in equilibrium, there is at most one link existing between a pair of users as shown in Statement (1). In the works [9][14] where content is perfectly substitutable, a user could choose to produce zero amount of contents in an equilibrium (like in a core-periphery structure) under the condition that he has already acquired a sufficient amount of contents from his friends. However, due to the heterogeneity of contents in our work, Statement (2) proves that a user always produces a positive amount of contents in equilibrium, since the contents he acquires from friends can never fully replace the contents produced by himself.

Lemma 2 provides a lower bound on the production level of an individual user at equilibrium. Due to the concavity of the benefit function and the linear production cost, the following lemma further provides an upper bound.

**Lemma 3.** *In any equilibrium* $\boldsymbol{s}^* = \left(\boldsymbol{x}^*, \boldsymbol{g}^*\right)$, $x_i^* \leq \overline{x}$ *for all* $i \in N$ *, where* $\overline{x}$ *is the unique solution of the equation* $v'\left(\overline{x}\right) = c$ *and is called as the maximum production of a user.*

*Proof*: See Appendix A. ■

Therefore, $\overline{x}$ serves as a constant upper-bound of the amount that an individual user is willing to produce in any equilibrium.

*B. Equilibrium in symmetric production profiles*

In this section, we study equilibria in symmetric production profiles. For the use of the analysis, we first present the related definitions of concepts.

**Definition 1 (Symmetric Production Profile).** A symmetric production profile $\left(\boldsymbol{x}, \boldsymbol{g}\right)$ satisfies the property $x_i = x_j,\ \forall i, j \in N$ .

Any strategy profile that is not a symmetric production profile belongs to asymmetric production profiles, which is correspondingly defined as follows.

**Definition 2 (Asymmetric Production Profile).** An asymmetric production profile $\left(\boldsymbol{x}, \boldsymbol{g}\right)$ satisfies the property $x_i \neq x_j,\ \exists i, j \in N$ .

The analysis of equilibria in asymmetric production profiles is postponed to the next section.

Correspondingly, we can also define a stronger version of symmetric production profile where each user does not only produce the same amount of contents, but also has the same degree (i.e. the same number of friends to share contents with).

**Definition 3 (Symmetric Profile).** *A symmetric profile* $\left(\boldsymbol{x}, \boldsymbol{g}\right)$ *is a symmetric production profile which satisfies the property* $\sum_{k \in N} \overline{g}_{ik} = \sum_{k \in N} \overline{g}_{jk},\ \forall i, j \in N$ .

Symmetric profiles constitute a subset of symmetric production profiles. Therefore, a symmetric equilibrium, which is a symmetric profile satisfying (6), is always a symmetric production equilibrium, which is a symmetric production profile satisfying (6). Nevertheless, we could also prove in the following proposition that all users' degrees are also the same in a symmetric production equilibrium and hence, a symmetric production equilibrium, on the other hand, is also a symmetric equilibrium.

**Proposition 1.** *In a symmetric production equilibrium* $\boldsymbol{s}^* = \left(\boldsymbol{x}^*, \boldsymbol{g}^*\right)$, $d_i\left(\overline{\boldsymbol{g}}^*\right) = d_j\left(\overline{\boldsymbol{g}}^*\right),\ \forall i, j \in N$ *always holds.*

*Proof*: See Appendix B. ■

Hence, in order to search for symmetric production equilibria, we only have to analyze symmetric profiles.

To facilitate our analysis, we define several concepts for a symmetric profile. First, we define an amount $z_i\left(d_i, d, x, c\right)$ as the optimal production level of a user $i$ when he has a degree $d_i$ and all the other users have a degree $d$ and a production level $x$ and the marginal cost of production is $c$. That is, $z_i\left(d_i, d, x, c\right)$ is the solution of

$$e\left(z_i\left(d_i, d, x, c\right), \left(z_i\left(d_i, d, x, c\right) + d_i x^\rho\right)^{\frac{1}{\rho}}\right) = c\,. \tag{8}$$

Additionally, we also define $\Delta r\left(d_i, d, x, c\right)$ as

$$\begin{aligned}\Delta r\left(d_i, d, x, c\right) \triangleq\ & v\left(\left[\left(z_i\left(d_i+1, d, x, c\right)\right)^\rho + \left(d_i+1\right)x^\rho\right]^{\frac{1}{\rho}}\right) - cz_i\left(d_i+1, d, x, c\right) \\ & - v\left(\left[\left(z_i\left(d_i, d, x, c\right)\right)^\rho + d_i x^\rho\right]^{\frac{1}{\rho}}\right) + cz_i\left(d_i, d, x, c\right)\end{aligned}\,. \tag{9}$$

It is obvious that two symmetric profiles having the same degree and production level provide users the same individual utility. Hence, these two symmetric profiles are regarded as being identical in realization and we have no preference on a particular one over the other. For illustration purposes, we use $\left(x, d\right)$ to denote a representative symmetric profile in which each user produces the amount $x$ and has the degree $d$.

Two conditions have to be satisfied for $\left(x, d\right)$ to be an equilibrium: (1) each user has no incentive to deviate from the amount $x$ in his production; (2) each user has no incentive to add or delete friends.

Condition (1) is satisfied when the marginal benefit of production equals to $c$, i.e. $e\left(x, X\right) = c$ where $X = \left(1+d\right)^{\frac{1}{\rho}} x$. This equation has a unique solution for each $d \in \left\{0, \ldots, n-1\right\}$, which is denoted as $x^s\left(d\right)$, and we also let $X^s\left(d\right) = \left(1+d\right)^{\frac{1}{\rho}} x^s\left(d\right)$. The following lemma characterizes $\left\{x^s\left(d\right)\right\}_{d=0}^{n-1}$ and $\left\{X^s\left(d\right)\right\}_{d=0}^{n-1}$.

**Lemma 4.** *$x^s\left(d\right)$ monotonically decreases with $d$, while $X^s\left(d\right) \triangleq \left(d+1\right)^{\frac{1}{\rho}} x^s\left(d\right)$ monotonically increases with $d$.*

*Proof*: See Appendix B. ■

To meet Condition (2), a user should have no incentive to deviate from his current subscription profile, which depends on the link cost $\gamma$. Particularly, if a user subscribes to some new friends, he receives some extra benefit by consuming the contents from these new friends. Nevertheless, the extra benefit should not be able compensate the user's extra link costs on the new subscriptions. Similarly, if a user deletes some existing links with his friends, the link costs saved in this way should not be able to cover the loss which it incurs from no longer receiving the contents from those deleted friends.

If a user $i$ deviates from a symmetric profile $\left(x^s\left(d\right),d\right)$ and changes his degree to $d_i \neq d$, his optimal production level becomes $z_i\left(d_i,d,x^s\left(d\right),c\right)$ and hence, $\sum_{l=d}^{d_i}\Delta r\left(l,d,x^s\left(d\right),c\right)$ represents the maximum gain (or minimum loss) on user $i$'s content utility if $d_i > d$ (or $d_i < d$).

The following proposition proves that $\Delta r\left(d_i,d,x^s\left(d\right),c\right)$ monotonically decreases against $d_i$. In other words, when a user deviates from a symmetric profile, the gain on his content utility from a new subscription monotonically decreases against his current degree. This is also consistent with the concavity of the benefit function (1).

**Proposition 2.** *Under a symmetric equilibrium with* $\left(x^s\left(d\right),d\right)$, $\Delta r\left(d_i,d,x^s\left(d\right),c\right)$ *monotonically decreases against* $d_i$ *for any* $i \in N$, $c > 0$ *and* $d \in \left\{0,\ldots,n-1\right\}$.

*Proof*: See Appendix B. ■

According to Proposition 2, if $\Delta r\left(d,d,x^s\left(d\right),c\right) < \gamma$, a user's utility does not increase by deviating from $d$ and adding a new subscription (i.e. his maximum gain on the content utility from the new subscription cannot cover the link cost $\gamma$). In this case, a user's utility does not increase by adding multiple subscriptions as well.

This analysis also works in the other direction. If $\Delta r\left(d-1,d,x^s\left(d\right),c\right) > \gamma$, a user's utility does not increase by deviating from $d$ and deleting an existing subscription (i.e. his minimum loss from the deleted subscription outweighs the link cost $\gamma$). Consequently, a user's utility does not increase by deleting multiple subscriptions as well.

Summarizing the above analysis, we can conclude that a symmetric profile $\left(x^s\left(d\right),d\right)$ is an equilibrium if and only if a user has no incentive to add or delete one subscription. This condition is formalized in the following theorem, which, in general, sums up the sufficient and necessary conditions for a symmetric production profile to be an equilibrium.

**Theorem 1.** *A symmetric production profile is an equilibrium if and only if the following conditions are satisfied:*

*(1) Each user has the same degree, denoted by* $d$*;*

*(2) Each user produces an amount* $x^s(d)$ *which is the solution of*

$$e\left(x^s(d),(1+d)^{\frac{1}{\rho}}x^s(d)\right)=c. \tag{10}$$

*(3) The following inequalities are satisfied*

$$\gamma>\underline{\gamma}(d)\triangleq\Delta r\left(d,d,x^s(d),c\right),\ \ if\ \ d\leq n-2, \tag{11}$$

*and*

$$\gamma<\overline{\gamma}(d)\triangleq\Delta r\left(d-1,d,x^s(d),c\right),\ \ if\ \ d\geq 1. \tag{12}$$

*Proof*: See Appendix B. ■

With the conditions in Theorem 1, we can prove the existence of symmetric equilibria in the following theorem. That is, for given $v$, $c$, and $\rho$, a non-empty region $\left(\underline{\gamma}(d),\overline{\gamma}(d)\right)$ always exists such that the symmetric profile $\left(x^s(d),d\right)$ is an equilibrium [5]. Moreover, for each value of $\gamma$, any two symmetric equilibria with different degrees cannot be equilibria at the same time.

**Theorem 2.** *(1) For given* $v$, $c$, *and* $\rho$, *there is a non-empty region* $\left(\underline{\gamma}(d),\overline{\gamma}(d)\right)$ *such that the symmetric profile* $\left(x^s(d),d\right)$ *is an equilibrium.*

*(2)* $\overline{\gamma}(d)<\underline{\gamma}(d-1)$ *always holds for all* $d\geq 1$.

*Proof*: See Appendix B. ■

We now look at whether our results are consistent with the results in [14] with $\rho\to 1$. When $\rho$ approaches 1, (10) becomes

[5] We define $\underline{\gamma}(n-1)=0$ and $\overline{\gamma}(0)=\infty$ for convenience.

$$\lim_{\rho\to 1} e\left(x^s\left(d\right),\left(1+d\right)^{\frac{1}{\rho}} x^s\left(d\right)\right) = v'\left(\left(1+d\right)x^s\left(d\right)\right) = c\,. \tag{13}$$

Hence, we have $\lim_{\rho\to 1} x^s\left(d\right) = \dfrac{\overline{x}}{1+d}$ . Substituting this into (11) and (12), we have that $\lim_{\rho\to 1} \underline{\gamma}\left(d\right) = \lim_{\rho\to 1} \overline{\gamma}\left(d\right) = \dfrac{c\overline{x}}{1+d}$. Hence, a symmetric profile $\left(x^s\left(d\right),d\right)$ with $d \in \left\{1,\ldots,n-2\right\}$ is a non-strict equilibrium if and only if $\gamma = \dfrac{c\overline{x}}{1+d}$. The only possible symmetric equilibria that could emerge are either with $d = 0$ , which form empty networks, or with $d = n - 1$, which form complete networks. This observation is the same as the results in [14], which verifies that our model is a generalization of that in [14].

We use the following exemplary benefit function

$$f\left(x_i, \boldsymbol{x}_{-i}, \boldsymbol{g}\right) = \log\left(1 + x_i^\rho + \sum_{j\in N_i\left(\overline{\boldsymbol{g}}\right)} x_j^\rho\right)^{\frac{1}{\rho}} \tag{14}$$

which satisfies Assumption (1) - (3) to illustrate $\left\{\left(\underline{\gamma}\left(d\right),\overline{\gamma}\left(d\right)\right)\right\}_{d=0}^{n-1}$ in Figure 1. It should be noted that we take the value $n = 10$ only for the illustration purpose. The results in Figure 1 can be extended to an arbitrarily large $n$ since symmetric equilibria are scalable. We take different values of $\rho$ , which represents different levels of users' desire for content diversity. Particularly, in Figure 1 (a), each segment at a degree $d$ represents the feasible region of $\gamma$ in which the symmetric profile $\left(x^s\left(d\right),d\right)$ is an equilibrium. As it shows, there is only a finite region of $\gamma$ where symmetric equilibria exist. Meanwhile, Figure 1 (b), which plots the average of the set $\left\{\overline{\gamma}\left(d\right)-\underline{\gamma}\left(d\right)\right\}_{d=1}^{n-2}$, illustrates the fact that the equilibrium region for each degree except 0 and n-1 diminishes as $\rho$ increases, i.e. when users have less desire for content diversity, which implies that to sustain a symmetric equilibrium becomes more difficult. When $\rho \to 1$, the feasible regions of $\gamma$ for all $d$ except $d = 0$ and $d = n-1$ disappear. This is consistent with the result in [14] as when contents become perfectly substitutable, there is no (strict) symmetric equilibrium except the following two: (1) the network is empty with no sharing between users, and each user produces the autarkic level $\overline{x}$ as defined in Lemma 2; and (2) a complete network where each user has the same production level and gets a total amount $\overline{x}$ of contents.

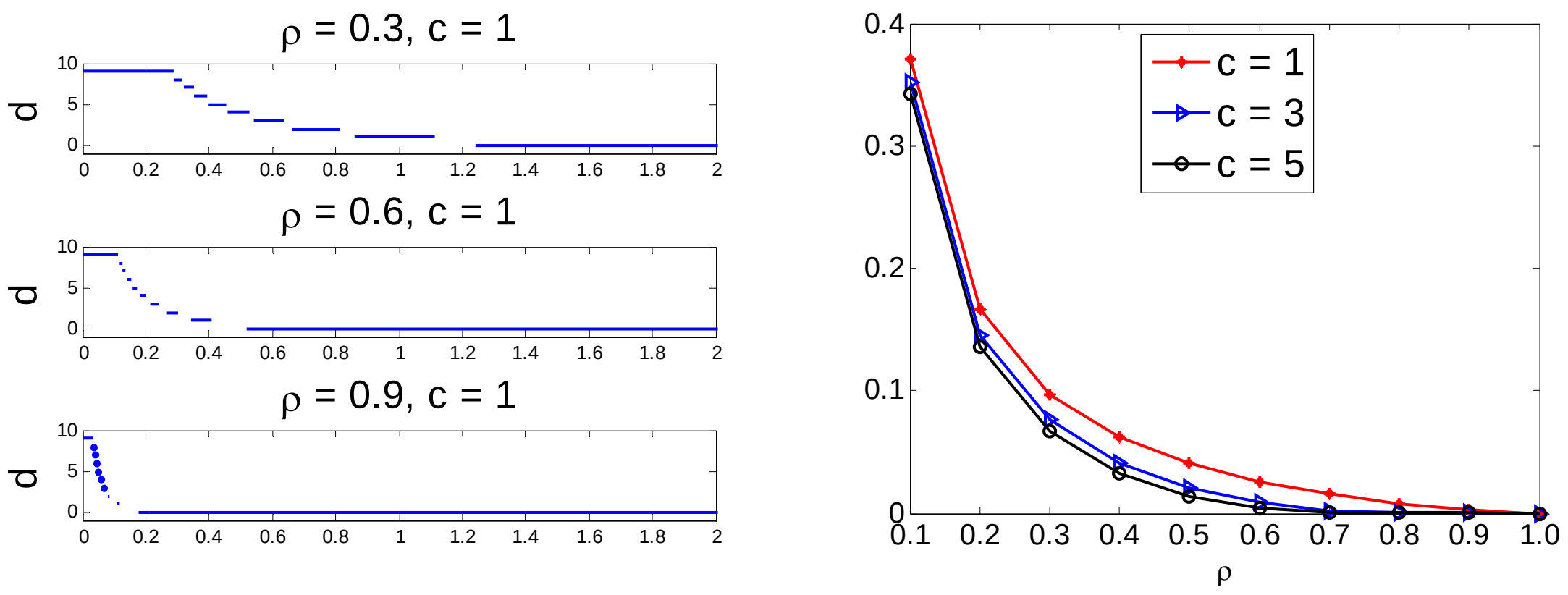

Figure 1 (a) The equilibrium region of $\gamma$ for $\left\{\left(x^s(d),d\right)\right\}_{d=0}^{n-1}$; (b) the average of $\left\{\overline{\gamma}(d)-\underline{\gamma}(d)\right\}_{d=1}^{n-2}$

*C. Equilibria in asymmetric production profiles*

In this section, we analyze the asymmetric production profiles as defined in Definition 2. An asymmetric equilibrium is thus defined as an asymmetric production profile that satisfies (6).

Without loss of generality, we order users by their production levels in an asymmetric production profile $\boldsymbol{s}=(\boldsymbol{x},\boldsymbol{g})$, i.e. $\overline{x}\geq x_1\geq x_2\geq\cdots\geq x_n$. Since not all users produce the same amount of contents in this case, there is always a positive number $n_h<n$, such that $x_i=x_{n_h}$ for all $i\leq n_h$. In the rest of this section, we call a user $i\leq n_h$ as a high producer and a user $j>n_h$ as a low producer. For notational convenience, we also let $\tilde{x}(\boldsymbol{x},\boldsymbol{g})$ denote the amount of contents produced by each high producer.

In Proposition 3, we characterize the asymmetric equilibria by summarizing their common properties. These properties are helpful in understanding the structure of an asymmetric equilibrium and facilitating the following analysis.

In this section, we first characterize the basic properties for an asymmetric equilibrium. Proposition 3 characterize the linking behavior of an asymmetric equilibrium.

**Proposition 3.** *In an asymmetric equilibrium* $\boldsymbol{s}^*=(\boldsymbol{x}^*,\boldsymbol{g}^*)$, *the following properties hold for* $\boldsymbol{g}^*$.

*(1)* $\overline{g}_{ij}^*=0,\ \exists i,j\in N$;

*(2) For each* $j>n_h$, $g_{ji}^*=1,\ \exists i\leq n_h$;

*(3)* $n_h\geq 2$ *always holds;*

*(4) For each* $i\leq n_h$, $g_{ij}^*=0,\ \forall j>n_h$;

*(5) For each* $i \le n_h$ , $\overline{g}_{ii'}^{*} = 0,\ \exists i' \le n_h$ *and* $i' \ne i$ .

*Proof*: See Appendix C. ■

Here we briefly explain Proposition 3. Statement (1) tells that there are at least two users in the network that are not friends with each other, which indicates that a network is always not a complete network in any asymmetric equilibrium. Statement (2) tells that each low producer subscribes to at least one high producer to acquire contents. Hence, users who produce more contents (e.g. celebrities and news agencies) are more favorable in an OSN. Statement (3) tells the fact that there does not exist an asymmetric equilibrium with a unique high producer. Therefore, the traditional star structure of content networks in which all users subscribe to a central news portal cannot be sustained. Statement (4) shows that a high producer do not subscribe any low producer but only acquire contents from other (if any) high producers. Hence, high producers play the role of influencers [13] in a network. Nevertheless, Statement (5) shows that each high producer does not connect with at least one other high producer in an asymmetric equilibrium. Therefore, high producers do not fully share contents among themselves, which leads to the content asymmetry among them.

Following Proposition 3, users' production behavior in an asymmetric equilibrium can be characterized in the following corollary.

**Corollary 1.** *In any asymmetric equilibrium* $\boldsymbol{s}^{*} = \left(\boldsymbol{x}^{*}, \boldsymbol{g}^{*}\right)$*, the following properties hold for* $\boldsymbol{x}^{*}$ *:*

*(1)* $\tilde{x}\left(\boldsymbol{x}^{*}, \boldsymbol{g}^{*}\right) < \overline{x}$ *always holds;*

*(2)* $\left|\left(X_i^{*}\right)^{\rho} - \left(X_j^{*}\right)^{\rho}\right| < \left(\tilde{x}\left(\boldsymbol{x}^{*}, \boldsymbol{g}^{*}\right)\right)^{\rho}$ .

*Proof:* See Appendix C. ■

Statement (1) tells that the highest amount of contents produced by an individual user in an asymmetric equilibrium is strictly smaller than $\overline{x}$ and thus the network cannot be empty in an asymmetric equilibrium. Statement (2) proves that the difference between any two users' perceived amounts of contents should be upper-bounded by $\tilde{x}\left(\boldsymbol{x}^{*}, \boldsymbol{g}^{*}\right)$, which is the highest amount of contents produced by an individual user in the equilibrium. Therefore, the distribution of accessible contents in the network should be sufficiently balanced among users in order to sustain an equilibrium. In particular topologies where accessible contents are highly unbalanced among users, no equilibrium can be sustained.

To better understand Proposition 3 and Corollary 1, we analyze the common asymmetric topologies in traditional content sharing networks such as the star and line, which are illustrated in Figure 2. By applying the results from Proposition 3 and Corollary 1, it can be prove that neither the star nor the line can be sustained in an equilibrium.

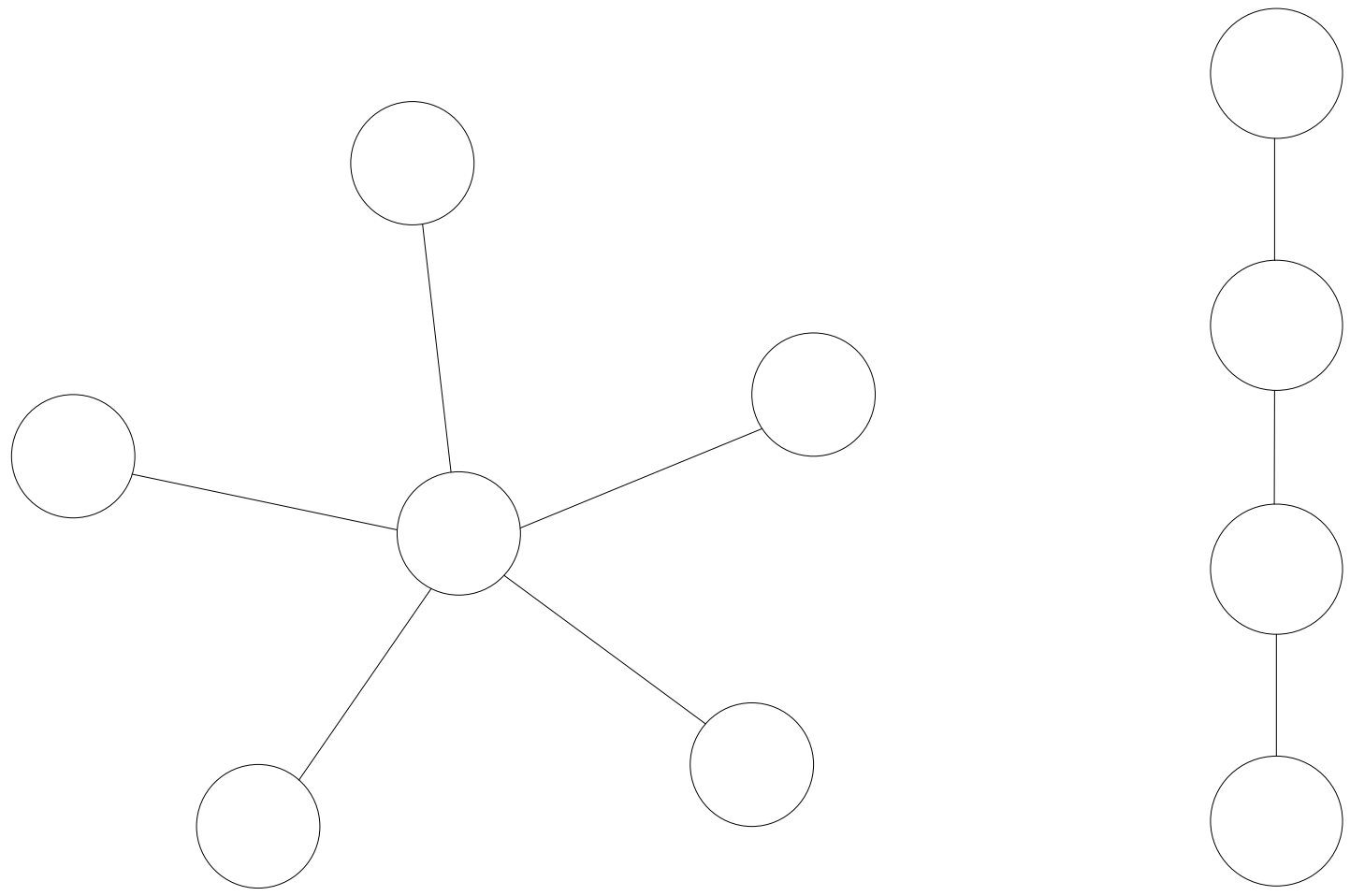

(a) (b)

Figure 2 (a) a star topology; (b) a line topology

**Corollary 2.** *(1) A star topology cannot emerge in any equilibrium when contents are not perfectly substitutable.*

*(2) A line topology cannot emerge in any equilibrium when contents are not perfectly substitutable.*

*Proof:* See Appendix C. ■

According to Proposition 3, high producers do not subscribe to low producers, whereas low producers always subscribe to some high producers. Hence, if a low producer does not subscribe to all high producers, he does not subscribe to any low producer as well. Summing up all these observations, we can conclude that there are three types of users in an asymmetric equilibrium, as summed up in the following proposition.

**Proposition 4.** *In any asymmetric equilibrium, each user belongs to one of the following three types:*

*(1) A high producer $i \le n_h$ who produces an amount $\tilde{x}\left(\boldsymbol{x}^*, \boldsymbol{g}^*\right)$ and does not subscribe to any low producer;*

*(2) A low producer $j > n_h$ who produces an amount $x_j^* < \tilde{x}\left(\boldsymbol{x}^*, \boldsymbol{g}^*\right)$ and subscribes to all high producers;*

*(3) A low producer $j > n_h$ who produces an amount $x_j^* < \tilde{x}\left(\boldsymbol{x}^*, \boldsymbol{g}^*\right)$ and subscribes to some high producers and does not sponsor links with any other low producer.*

*Proof:* We have shown in Proposition 3 (4) and (5) the subscription behavior of high producers in an asymmetric equilibrium. For a low producer $j$ who does not subscribe to any other low producer, he belongs to type (3). For a low producer $j$ who subscribes to another low producer $j'$, it is obvious that $j$ subscribes to all high producers. Otherwise, $j$ could strictly increase his utility by switching his subscription from $j'$ to a high producer $i \leq n_h$ who he does not subscribe to. ■

Proposition 4 classifies users in an asymmetric equilibrium and characterizes their subscription behaviors, respectively. Depending on the production level of a high producer, i.e. the value of $\tilde{x}\left(\boldsymbol{x}^*, \boldsymbol{g}^*\right)$, there could be multiple equilibria under the same network conditions. Figure 3 illustrates example equilibria in a network of $n = 9$ user when $\rho = 0.8$, $c = 0.1$, and $\gamma = 2$. As it shows, different equilibria with different production levels can emerge. In Figure 3 (a), there are four high producers (marked red) with $\tilde{x}\left(\boldsymbol{x}^*, \boldsymbol{g}^*\right)$ being large. As each low producer (marked green) can get sufficient contents from high producers, he does not subscribe to any other low producer. In this case, there are only two levels of content productions in the network. However in Figure 3 (b), with less (two) high producer (marked red) and a smaller value of $\tilde{x}\left(\boldsymbol{x}^*, \boldsymbol{g}^*\right)$, each low producer cannot get sufficient contents from high producers and he also subscribes to other low producers, which leads to an equilibrium with three levels of content productions, where each user in green produces a medium amount of contents and each user in blue has the lowest production level and subscribes to all users whose production levels are higher than him.

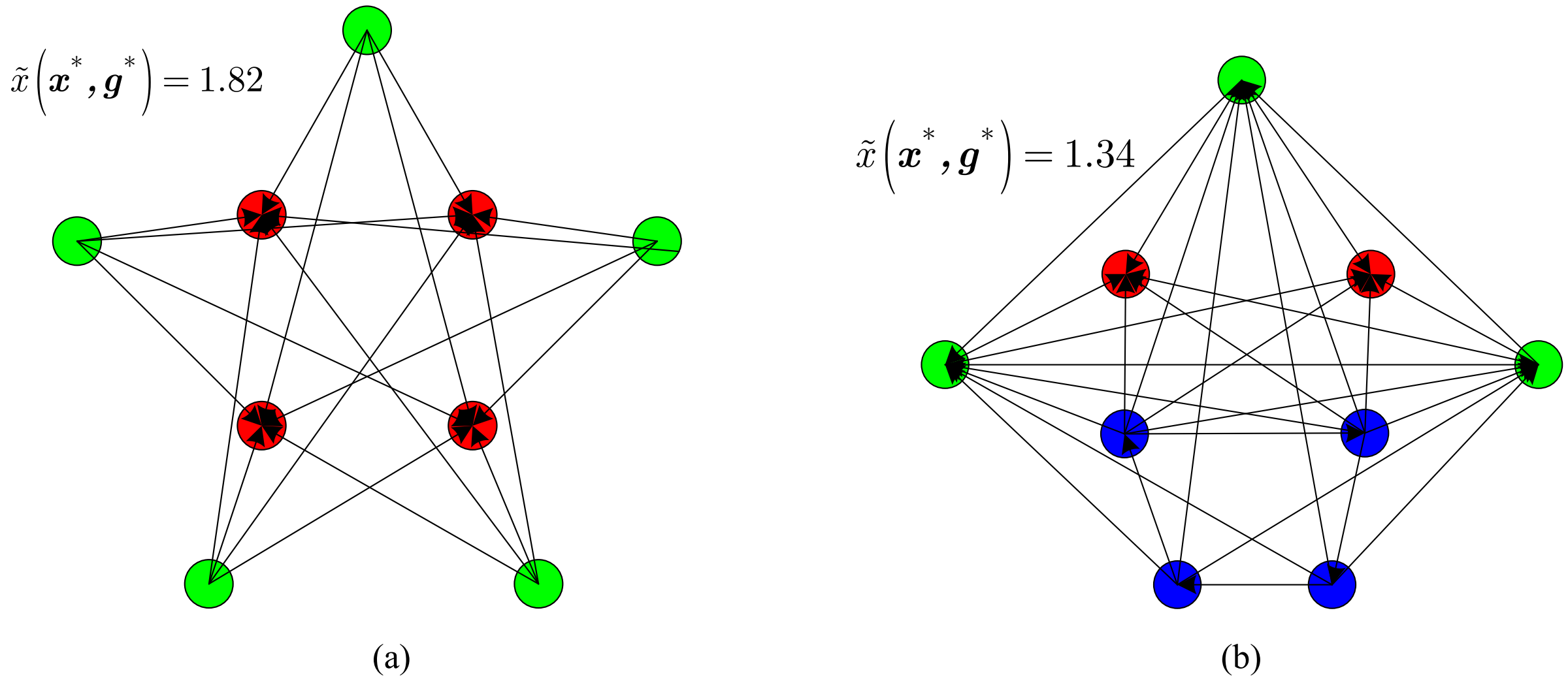

Figure 3 Two examplary asymmetric equilibria: (a) two-level production; (b) three level production

In contrast to the multiple production levels exhibited in Figure 3, we prove in the following theorem, by taking the limit on the size of a network to infinity, that in an asymmetric equilibrium $\left(\boldsymbol{x}^*, \boldsymbol{g}^*\right)$, there are only two types of users in the network depending on their production levels. Each high producer produces an amount $\tilde{x}\left(\boldsymbol{x}^*, \boldsymbol{g}^*\right)$ and subscribes to $\tilde{k}\left(\boldsymbol{x}^*, \boldsymbol{g}^*\right)$ other high producers. Each low producer produces an amount $\underset{\sim}{x}\left(\boldsymbol{x}^*, \boldsymbol{g}^*\right)$ and subscribes to $\underset{\sim}{k}\left(\boldsymbol{x}^*, \boldsymbol{g}^*\right)$ high producers. Meanwhile, each high producer does not subscribe to any low producers, and low producers do not mutually subscribe to each other. Therefore, all subscriptions in the network lead to high producers, which can be regarded as the major sources of contents in the network. The network then exhibits a flat structure with only two levels in its user hierarchy. We call the high producers as influencers and the low producers as subscribers whose main goal is to acquire contents from influencers.

**Theorem 3.** When the network population $n$ is sufficiently large, only two types of users exist in any asymmetric equilibrium $\left(\boldsymbol{x}^*, \boldsymbol{g}^*\right)$: an influencer who produces an amount $\tilde{x}\left(\boldsymbol{x}^*, \boldsymbol{g}^*\right)$ and subscribes to $\tilde{k}\left(\boldsymbol{x}^*, \boldsymbol{g}^*\right)$ other influencers, and a subscriber who produces an amount $\underset{\sim}{x}\left(\boldsymbol{x}^*, \boldsymbol{g}^*\right)$ and subscribes to $\underset{\sim}{k}\left(\boldsymbol{x}^*, \boldsymbol{g}^*\right)$ influencers. The production and subscription profiles have the following relationship as $\tilde{x}\left(\boldsymbol{x}^*, \boldsymbol{g}^*\right) > \underset{\sim}{x}\left(\boldsymbol{x}^*, \boldsymbol{g}^*\right)$ and $\tilde{k}\left(\boldsymbol{x}^*, \boldsymbol{g}^*\right) < \underset{\sim}{k}\left(\boldsymbol{x}^*, \boldsymbol{g}^*\right)$.

*Proof*: See Appendix C. ■

A key observation in the proof of Theorem 3 is that due to the cost of subscriptions, a user has to produce a sufficiently large amount of contents, which is lower-bounded away from 0, in order to attract others to subscribe to him. Meanwhile, a user has an upper bound on the amount of contents he would like to consume due to the concavity of the benefit function (1). As a result, a user also has an upper bound on the number of subscriptions he would like to maintain. Now as $n \to \infty$, the number of influencers also goes to infinity which provides enough contents for each user in the network to subscribe with. Hence, low producers will not mutually subscribe to each other but only subscribe to high producers.

Theorem 3 proves that the two-level structure is a necessary condition for an asymmetric equilibrium when $n$ is sufficiently large, which is shown in Figure 4 with two examples of equilibrium topologies. Both topologies exhibit the two-level structure composed of two rings. The inner ring represents the influencers that produce a large amount of contents, and the outer ring represents the subscribers that produce a small amount of contents. Influencers compose the center of the network, they do not subscribe to any subscriber, but they will mutually subscribe to each other in order to share contents among themselves. Subscribers, on the contrary, only subscribe to influencers but not to any other subscribers. Hence, our results are consistent with the empirical observations that a majority of users in an OSN get most of their contents from a relatively small group of high producers. Also recall that there is only a limited region of $\gamma$ where symmetric equilibria exist and hence, asymmetric equilibria with the two-level structure are more likely to emerge in OSNs.

With the link cost increases, as shown in Figure 4 when $\gamma$ changes from 2 to 5, the network becomes sparse as in Figure 4 (b). Instead of subscribing to multiple influencers, subscribers only get contents from influencers from their vicinity and the network is divided into many small sub-networks where each influencer takes the charge of providing contents to his local users.

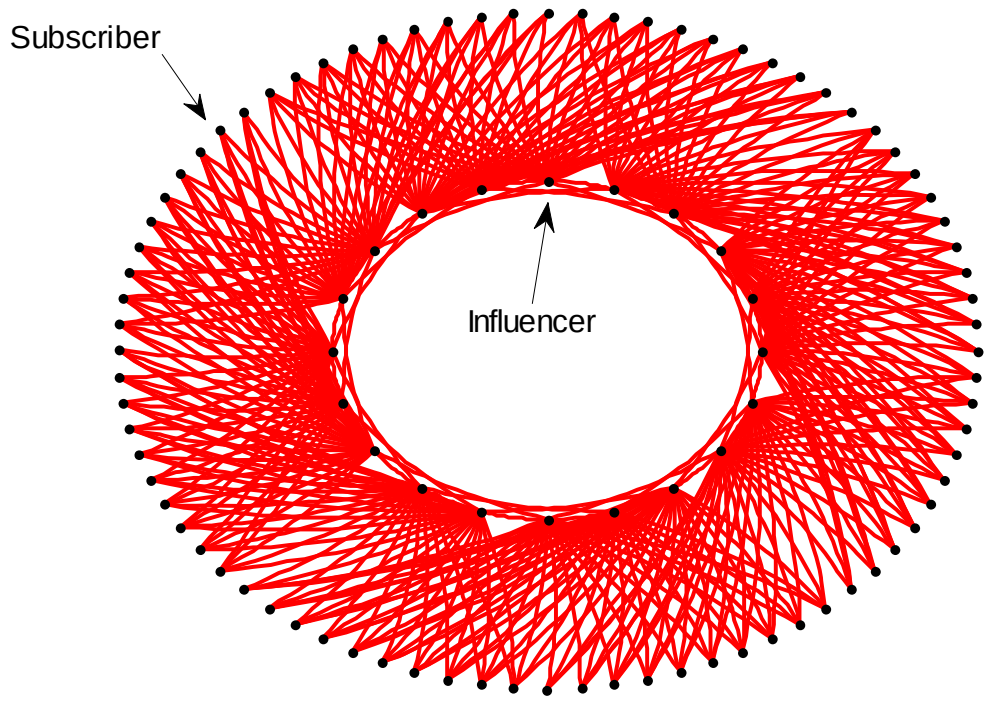

(a) $\gamma = 2$

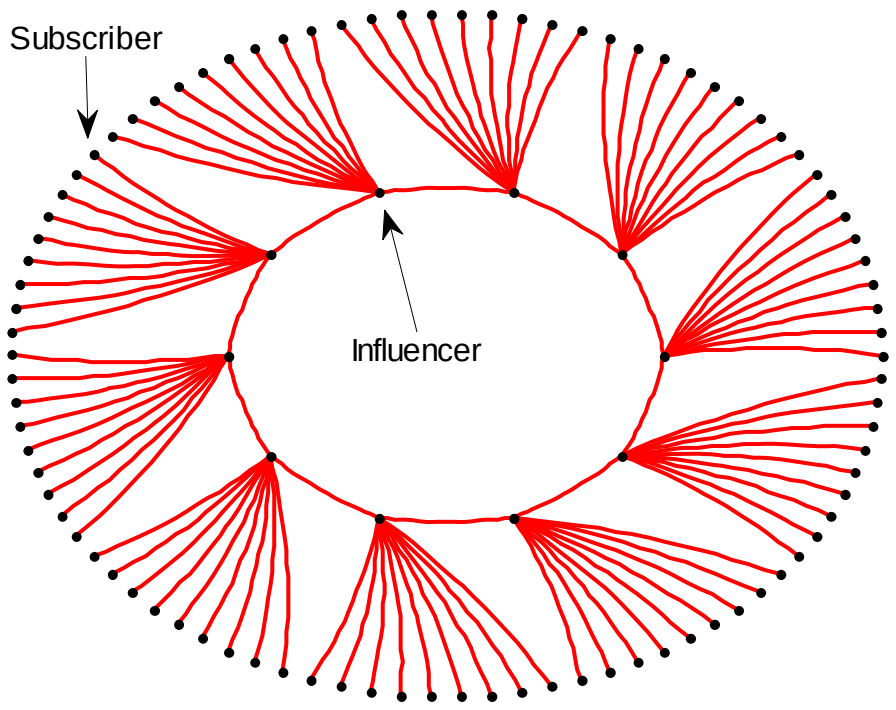

(b) $\gamma = 5$

Figure 4 Examplary equilibria in an OSN with $n = 100$ users.

The influencer-subscriber structure exhibits some similarity to the small world networks. Nevertheless, different from the law of the few in [14] where there are a fixed number of influencers whose population size is independent on the network size. We further prove in the following theorem that the fraction of influencers in the population does not go to zero as $n$ tends to infinity. This indicates that more influencers will emerge with the growth of the network, which is due to the content heterogeneity. As more users join the network, the content production which is monopolized by a small number of powerful influencers can no longer satisfy users' desire for diverse contents. As a result, new influencers with different varieties of contents should emerge to attract users and stabilize the network.

**Theorem 4.** The law of the few does not hold when contents are not perfectly substitutable, as there exists an $\varepsilon > 0$ such that in any equilibrium

$$\lim_{n \to \infty} \frac{n_h}{n} > \varepsilon \,. \tag{15}$$

*Proof*: See Appendix C. ■

Figure 5 plots how the fraction of influencers and users' production levels change in asymmetric equilibria as the population size $n$ increases [6]. Figure 5 (a) verifies Theorem 4. The number of influencers grows at a slower speed than $n$ at the beginning and hence, $\frac{n_h}{n}$ monotonically decreases. However, when $n$ is sufficiently large, the number of subscribers that an influencer can support reaches its upper bound as discussed in Theorem 3. Hence, to support an equilibrium, more influencers will emerge and $\frac{n_h}{n}$ stops decreasing, which indicates that the number of influencers starts to grow proportionally with $n$. Figure 5 (b) shows how the production levels of influencers and subscribers change against $n$. For a better illustration, we plot the normalized production levels which are compared with their values at $n = 100$ on each curve. As it shows, the normalized production level of subscribers drops more drastically than that of influencers as the network size grows, since a subscriber mainly relies on the contents acquired from others rather than self-production.

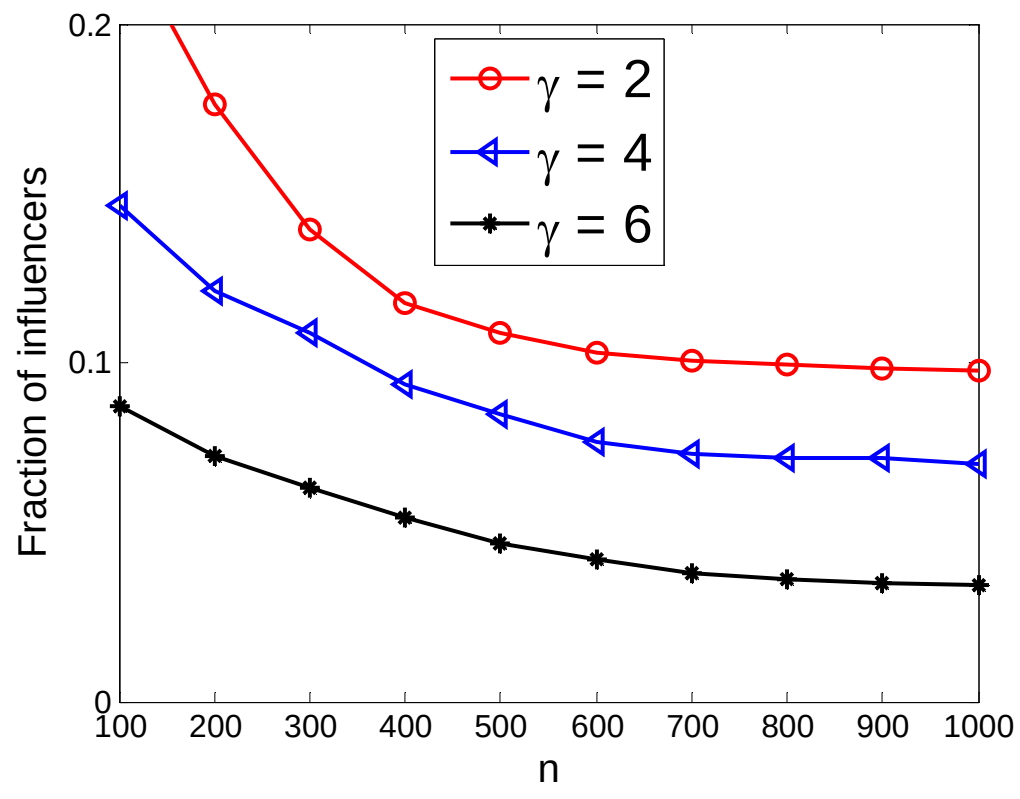

(a)

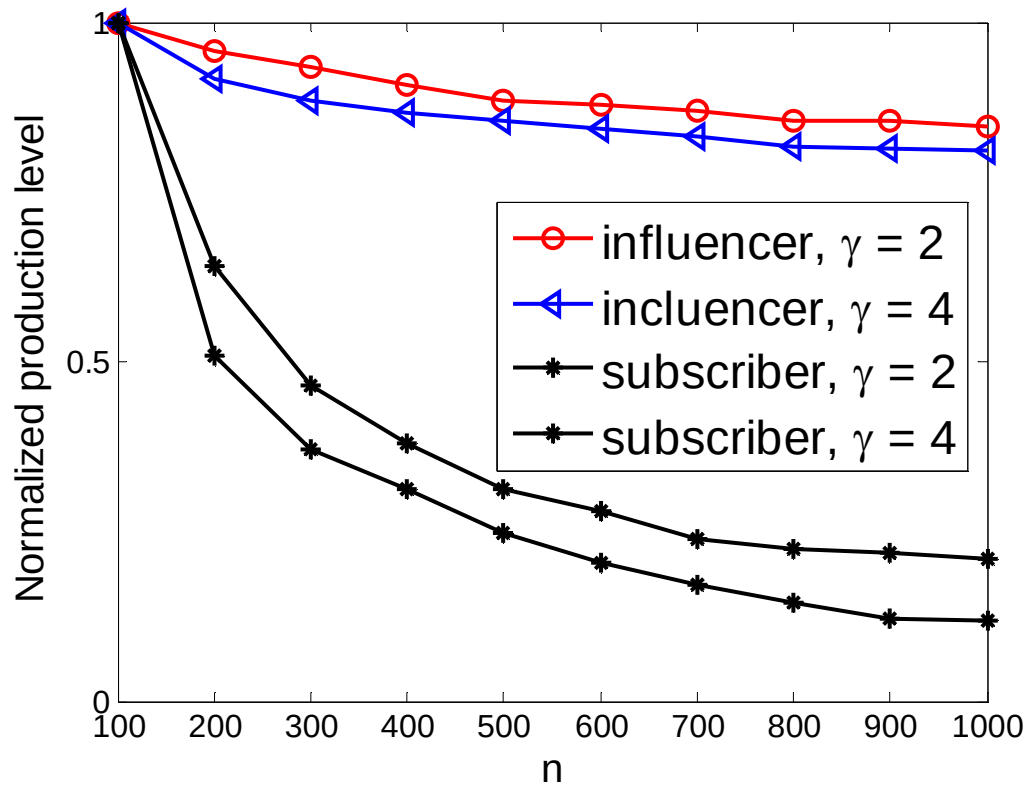

(b)

Figure 5 (a) The fraction of influencers changing against $n$;
(b) The normalized production levels of users changing against $n$.

Before ending this section, we further provide some characterization on the influencer-subscriber structure.

**Corollary 3.** *(1)In the influencer-subscriber structure, the utility received by an influencer is always higher than the utility received by a subscriber.*

*(2) An influencer always maintains fewer subscriptions than a subscriber.*

*Proof:* See Appendix C. ■

[6] It should be noted that for a given set of the network parameters $\gamma$, $c$, $\rho$, and $v$, there are multiple asymmetric equilibria. Hence, we use the monte-carlo method to run the experiments multiple times for each value of $n$ and plot the average results in Figure 3.

# IV. Social Optimum and Pricing Scheme

We now study the social optimum of the PNF game. The social welfare of a network is defined to be the sum of users' individual utilities. For a profile $\left(\boldsymbol{x}, \boldsymbol{g}\right)$, the social welfare is given by $W\left(\boldsymbol{x}, \boldsymbol{g}\right) = \sum_{i \in N} u_i\left(\boldsymbol{x}, \boldsymbol{g}\right)$. A profile $\left(\boldsymbol{x}^{\#}, \boldsymbol{g}^{\#}\right)$ achieves the social optimum if

$$W\left(\boldsymbol{x}^{\#}, \boldsymbol{g}^{\#}\right) \geq W\left(\boldsymbol{x}, \boldsymbol{g}\right), \forall \left(\boldsymbol{x}, \boldsymbol{g}\right). \tag{16}$$

First, we prove that the social optimum can be achieved by a symmetric profile.

**Theorem 5.** The social optimum can be achieved by a symmetric profile.

*Proof*: See Appendix D. ■

Using the same notations as in Section III.B, we let $\left(x, d\right)$ to refer a symmetric profile, where the social welfare can be represented as follows:

$$W\left(x, d\right) = \sum_{i \in N} u_i\left(x, d\right) = n\left(v\left(\left(1+d\right)^{\frac{1}{\rho}} x\right) - cx - \frac{d}{2}\gamma\right), \tag{17}$$

The social optimum is the solution of the following problem:

$$\begin{aligned} &\max_{x,d} \left\{ v\left(\left(1+d\right)^{\frac{1}{\rho}} x\right) - cx - \frac{d}{2}\gamma \right\} \\ &s.t. \quad x \geq 0 \\ &\qquad\ \ d \in \left\{0, 1, \ldots, n-1\right\} \end{aligned}. \tag{18}$$

which can be solved sequentially by two steps. In the first step, we derive the optimal production level for each $d$, denoted as $x^{\#}\left(d\right)$. It is the solution of the following equation:

$$\left(1+d\right)^{\frac{1}{\rho}} v'\left(\left(1+d\right)^{\frac{1}{\rho}} x^{\#}\left(d\right)\right) = c, \ \forall d \in \left\{0, \ldots, n-1\right\}. \tag{19}$$

In the second step, we optimize over all symmetric profiles $\left\{\left(x^{\#}\left(d\right), d\right)\right\}_{d=0}^{n-1}$ to find the optimal degree $d^{\#}$. $\left(x^{\#}\left(d^{\#}\right), d^{\#}\right)$ then constitutes the optimal strategy profile that achieves the social optimum.

Denote a user's content utility as follows for convenience

$$q\left(x^{\#}\left(d\right), d\right) \triangleq v\left(\left(1+d\right)^{\frac{1}{\rho}} x^{\#}\left(d\right)\right) - cx^{\#}\left(d\right), \tag{20}$$

we prove that as $d$ increases, the change on $q\left(x^{\#}\left(d\right),d\right)$, i.e.

$$\Delta q\left(x^{\#}\left(d\right),d\right) \triangleq q\left(x^{\#}\left(d+1\right),d+1\right) - q\left(x^{\#}\left(d\right),d\right) \tag{21}$$

monotonically decreases.

**Proposition 5.** $\Delta q\left(x^{\#}\left(d\right),d\right)$ that is defined in (21) decreases with $d$.

*Proof*: See Appendix D. ■

By changing the degree from $d$ to $d+1$, each user's content utility increases by $\Delta q\left(x^{\#}\left(d\right),d\right)$ and he pays an additional link cost of $\frac{\gamma}{2}$ on average. According to Proposition 5, the optimal degree $d^{\#}$ is determined as in the following proposition.

**Proposition 6.** The optimal degree $d^{\#}$ satisfies the following inequalities:

$$\begin{aligned} &\Delta q\left(x^{\#}\left(d\right),d\right) < \frac{\gamma}{2},\ \ if\ \ d^{\#} < n-1, \\ &\Delta q\left(x^{\#}\left(d-1\right),d-1\right) > \frac{\gamma}{2},\ \ if\ \ d^{\#} > 0. \end{aligned} \tag{22}$$

*Proof*: Similar to the proof of Theorem 1 and is omitted here. ■

In order to sustain $\left(x^{\#}\left(d^{\#}\right),d^{\#}\right)$ as an equilibrium, we propose a pricing scheme where users have to pay extra fees for subscriptions and the acquired contents from their friends. The pricing scheme is flat-rate such that each user pays a constant fee $t$ for each subscription and a constant fee $p$ for unit content acquired. It should be noted that different from $\gamma$ that is paid to the service provider of the network, both $t$ and $p$ are paid directly to the user who is being subscribed. Meanwhile, $t$ and $p$ are not necessarily positive. If $t$ or $p$ is negative, it means that a user should be granted with some reward for creating subscriptions in order to improve the content sharing efficiency in the network. With the pricing scheme, a user $i$'s utility becomes

$$\begin{aligned} \tilde{u}_i\left(\boldsymbol{x},\boldsymbol{g}\right) = v\left(\left[x_i^{\rho} + \sum_{j\in N_i\left(\overline{\boldsymbol{g}}\right)} x_j^{\rho}\right]^{\frac{1}{\rho}}\right) - cx_i - p\sum_{j\in N_i\left(\overline{\boldsymbol{g}}\right)} x_j \\ +px_i\left|N_i\left(\overline{\boldsymbol{g}}\right)\right| - \left(\gamma+t\right)\left|N_i\left(\boldsymbol{g}\right)\right| + t\left|N_i\left(\overline{\boldsymbol{g}}\right)/N_i\left(\boldsymbol{g}\right)\right| \end{aligned} \tag{23}$$

Recall that the purpose of the pricing scheme is to sustain $\left(x^{\#}\left(d^{\#}\right), d^{\#}\right)$ to be an equilibrium, we have to ensure that the user's marginal benefit of production equals to $c$, i.e.

$$\left(1+d^{\#}\right)^{\frac{1-\rho}{\rho}} v'\left(\left(1+d^{\#}\right)^{\frac{1}{\rho}} x^{\#}\left(d^{\#}\right)\right) + pd^{\#} = c \tag{24}$$

Meanwhile, in order to enforce a user to maintain $d^{\#}$ friends, we have to ensure that the maximum gain (minimum loss) on the content utility that a user of the degree $d^{\#}$ can obtain by adding (deleting) a subscription is smaller (larger) than the related cost, i.e. $\left(\gamma+t\right)+px^{\#}\left(d^{\#}\right)$. The desirable values of $p$ and $t$ are obtained in the following theorem.

**Theorem 6.** The social optimum in a non-cooperative equilibrium if the prices are set to be $p^{\#}$ and $t^{\#}$ as follows:

$$p^{\#} = \frac{d^{\#}}{1+d^{\#}} c, \tag{25}$$

$$\begin{aligned} t^{\#} > \Delta r\left(d^{\#}, d^{\#}, x^{\#}\left(d^{\#}\right), c - p^{\#}d^{\#}\right) \\ - p^{\#}x^{\#}\left(d^{\#}\right) - \gamma \end{aligned}, \; if \;\; d^{\#} \leq n\text{ - }2, \tag{26}$$

and

$$\begin{aligned} t^{\#} < \Delta r\left(d^{\#}-1, d^{\#}, x^{\#}\left(d^{\#}\right), c - p^{\#}d^{\#}\right) \\ - p^{\#}x^{\#}\left(d^{\#}\right) - \gamma \end{aligned}, \; if \;\; d^{\#} \geq 1., \tag{27}$$

*Proof*: Similar to the proof of Theorem 1 and is omitted here. ■

Figure 6 plots how the optimal prices change against the link cost $\gamma$ [7]. When $\gamma$ is small, users prefer to acquire contents from others rather than self-production, due to the low cost of content transmission. At this stage, high prices for subscriptions and content acquisitions should be set in order to increase users' production levels. Nevertheless, as $\gamma$ increases, users tend to produce contents by themselves and the content sharing level in the network becomes low. In this case, both $p$ and $t$ should be reduced in order to encourage content sharing. As Figure 6 (b) shows, when $\gamma$ is too large, $t^{\#}$ becomes negative, which implies that a user should be compensated in his subscription to cover his link cost.

---

[7] Since there is a range of applicable $t^{\#}$ for each value of $\gamma$ according to **Error! Reference source not found.**, we take the median of this range in the plot.

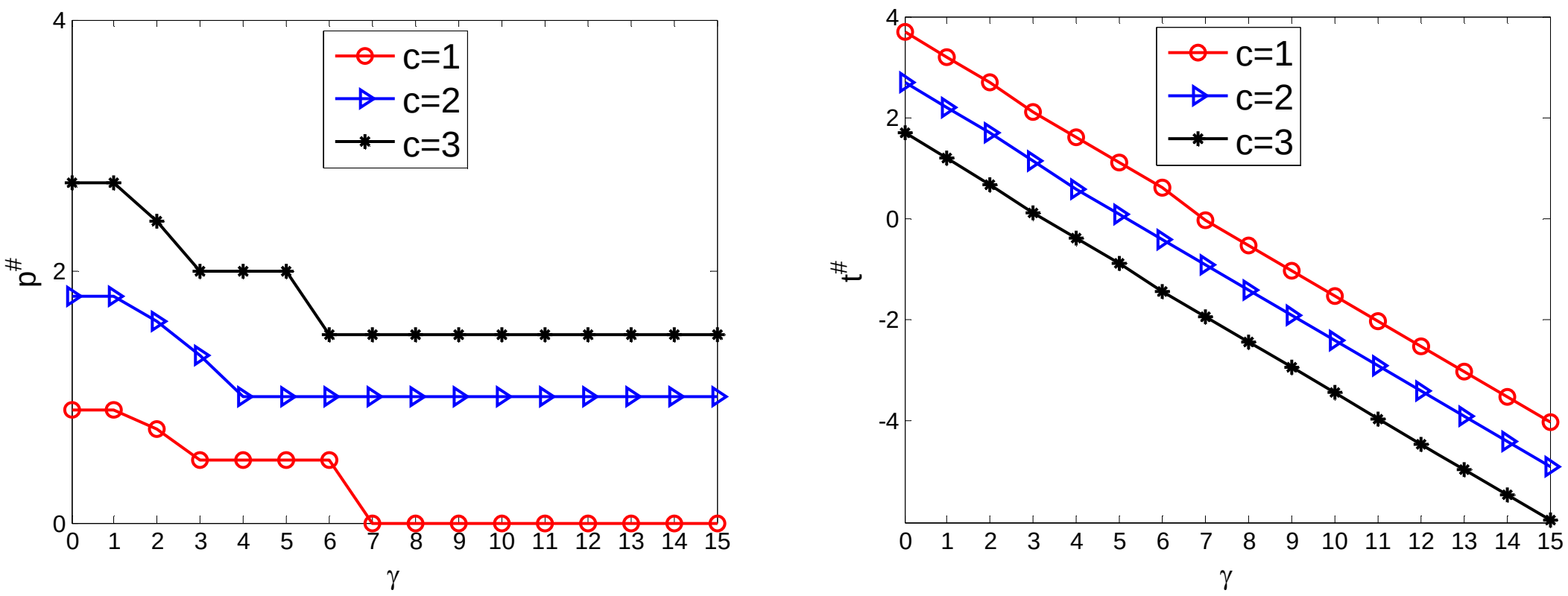

(a) (b)

Figure 6 The optimal prices $p^{\#}$ and $t^{\#}$ changing against $\gamma$ .

# V. Conclusion and Future Research

In this paper, we investigate the problem of network formation in OSNs. Different from the existing literature, the users' incentives for producing contents themselves and for creating links to consume the contents produced by other are jointly considered. Moreover, we determine rigorously how the users' desire for heterogeneous contents impacts on the users' interactions. Using a game-theoretic analysis, we show that as the size of the OSN grows, every (strict) non-cooperative equilibrium is composed of either a symmetric topology or a two-level hierarchical topology where the number of influencers grows proportionally with the network size. The social optimum can be achieved in a symmetric topology, and we design a pricing scheme to achieve it in a non-cooperative equilibrium. Our analysis can be extended in several directions, among which we mention two. First, users in an OSN do not necessarily need to be homogeneous as discussed in this paper. Different users might perceive contents differently by having different benefit functions. Second, alternative models on content transmission and subscription can be adopted. For instance, user subscriptions are free on Twitter while the content transmission on an established link is unilateral (i.e. a user being followed cannot acquire contents shared by his followers.). Moreover, network formation with indirect content transmission also forms an important future direction, where content of users who are not directly connected can also be shared.

# Appendix A

*1) Proof of Lemma 2*

(1) Suppose that $\boldsymbol{s}^* = \left(\boldsymbol{x}^*, \boldsymbol{g}^*\right)$ with $g_{ij}^* g_{ji}^* = 1$, then user $i$ can strictly increase his utility by selecting $g_{ij} = 0$, which leads to a contradiction to the definition of an equilibrium. Hence, Statement (1) follows.

(2) When user $i$ is isolated, i.e. $\left|N_i\left(\overline{\boldsymbol{g}}\right)\right| = 0$, the first-order derivative in $x_i$ over his benefit function becomes $\frac{\partial f}{\partial x_i} = v'\left(x_i\right)$. Since $v'\left(0\right) \geq \alpha > c$, user $i$ always has the incentive to produce at $x_i = 0$.

When user $i$ is not isolated, i.e. $\left|N_i\left(\overline{\boldsymbol{g}}\right)\right| > 0$, we have $\lim_{x_i \to 0} \frac{\partial f}{\partial x_i} = v'\left(x_i\right)\left(\frac{X_i}{x_i}\right)^{1-\rho} \to \infty$ and hence user $i$ also has the incentive to produce at $x_i = 0$. Summing up, we have Statement (2) follows. ■

*2) Proof of Lemma 3*

Due to Assumption (3), user $i$'s marginal benefit of production monotonically decreases with the amount of contents he acquires from others. Hence, user $i$ has the largest marginal benefit of production, i.e. the largest incentive to produce content, at every point of $x_i$ when he acquires no content from others, i.e. $\left|N_i\left(\overline{g}\right)\right| = 0$. The corresponding utility function for user $i$ can be rewritten as

$$u_i\left(\boldsymbol{x}, \boldsymbol{g}\right) = v\left(x_i\right) - cx_i. \tag{28}$$

He stops producing new contents when the marginal benefit of production equals to the marginal cost, i.e. at the point $x_i = \overline{x}$ where $v'\left(\overline{x}\right) = c$. Since $v\left(x\right)$ is strictly concave, $v'\left(x\right)$ is strictly decreasing and hence, $v'\left(\overline{x}\right) = c$ has a unique solution. ■

## Appendix B

*3) Proof of Proposition 1*

Suppose each user produces the amount $x^*$ and there are two different users $i$ and $j$ with $d_i\left(\overline{\boldsymbol{g}}^*\right) > d_j\left(\overline{\boldsymbol{g}}^*\right)$ without loss of generality. In a symmetric equilibrium, both user $i$ and $j$ have no incentive to produce more contents than $x^*$. Hence, it could be determined from (3) that the marginal benefits of content production for $i$ and $j$ at $x^*$ should both equal to $c$, i.e.

$$e\left(x^*, X_i^*\right) = e\left(x^*, X_j^*\right) = c. \tag{29}$$

However, according to Assumption (3), we should have $e\left(x^*, X_i^*\right) < e\left(x^*, X_j^*\right)$ since user $i$ acquires more contents from others than user $j$ and hence, there is a contradiction and the assumption that $d_i\left(\overline{\boldsymbol{g}}^*\right) > d_j\left(\overline{\boldsymbol{g}}^*\right)$ does not hold in a symmetric production equilibrium. Therefore, Proposition 1 follows. ■

*4) Proof of Lemma 4*

Using the equilibrium condition, we have that

$$e\left(x^s\left(d\right), X^s\left(d\right)\right) = v'\left(\left(d+1\right)^{\frac{1}{\rho}} x^s\left(d\right)\right)\left(d+1\right)^{1-\rho} = c\,. \tag{30}$$

Now consider an arbitrary value of $d$. According to Assumption (3), the following inequality holds:

$$v'\left(\left(d+2\right)^{\frac{1}{\rho}} x^s\left(d\right)\right)\left(d+2\right)^{1-\rho} < v'\left(\left(d+1\right)^{\frac{1}{\rho}} x^s\left(d\right)\right)\left(d+1\right)^{1-\rho}\,. \tag{31}$$

Regarding the fact that $v'\left(\left(d+2\right)^{\frac{1}{\rho}} x^s\left(d+1\right)\right)\left(d+2\right)^{1-\rho} = v'\left(\left(d+1\right)^{\frac{1}{\rho}} x^s\left(d\right)\right)\left(d+1\right)^{1-\rho}$ and $v\left(\bullet\right)$ is strictly concave, we have $x^s\left(d+1\right) < x^s\left(d\right)$. As $d$ is arbitrarily chosen, it can be concluded that $x^s\left(d\right)$ monotonically decreases with $d$.

Similarly, regarding the fact that $v'\left(X^s\left(d+1\right)\right)\left(d+2\right)^{1-\rho} = v'\left(X^s\left(d\right)\right)\left(d+1\right)^{1-\rho}$, we have $v'\left(X^s\left(d+1\right)\right) < v'\left(X^s\left(d\right)\right)$ and thus $X^s\left(d+1\right) > X^s\left(d\right)$. Therefore, $X^s\left(d\right)$ monotonically increases with $d$. ■

*5) Proof of Proposition 2*

Taking the derivative in $d_i$, we have that

$$\frac{\partial \Delta r\left(d_i, d, x^s\left(d\right), c\right)}{\partial d_i}$$

$$= \frac{\partial\left[v\left(\left[\left(z_i\left(d_i+1, d, x, c\right)\right)^\rho + \left(d_i+1\right)x^\rho\right]^{\frac{1}{\rho}}\right)\right]}{\partial d_i} - \frac{\partial\left[v\left(\left[\left(z_i\left(d_i, d, x, c\right)\right)^\rho + d_i x^\rho\right]^{\frac{1}{\rho}}\right)\right]}{\partial d_i} . \tag{32}$$
$$- c\left(\frac{\partial z_i\left(d_i+1, d, x, c\right)}{\partial d_i} - \frac{\partial z_i\left(d_i, d, x, c\right)}{\partial d_i}\right)$$
$$= \frac{c}{\rho\left(z_i\left(d_i+1, d, x, c\right)\right)^{\rho-1}}x^\rho - \frac{c}{\rho\left(z_i\left(d_i, d, x, c\right)\right)^{\rho-1}}x^\rho$$

It should be noted that the second inequality of (32) relies on the fact that

$$\left.\frac{\partial\left[v\left(\left[x_i^\rho + d_i x^\rho\right]^{\frac{1}{\rho}}\right)\right]}{\partial x_i}\right|_{x_i = z_i\left(d_i, d, x, c\right)} = c\,. \tag{33}$$

Using a similar argument as Lemma 4, we have that $z_i\left(d_i+1, d, x, c\right) < z_i\left(d_i, d, x, c\right)$ always holds and hence $\frac{\partial \Delta r\left(d_i, d, x^s\left(d\right), c\right)}{\partial d_i} < 0$. Therefore, Proposition 2 follows. ■

*6) Proof of Theorem 1*

Statements (1) and (2) are obvious and hence, we only prove Statement (3) here.

First we consider the case that $d \le n - 2$. For a user $i$, if $\Delta r\left(d, d, x^s\left(d\right), c\right) < \gamma$, then we have $\Delta r\left(d_i, d, x^s\left(d\right), c\right) < \gamma$ for all $d_i > d$. Suppose user $i$ add $\Delta d_i \ge 1$ subscriptions, the maximum gain on his content utility is $\sum_{l=0}^{\Delta d_i - 1} \Delta r\left(d+l, d, x^s\left(d\right), c\right)$, which is smaller than $\Delta d_i \gamma$. As a result, no user will have the incentive to add new subscriptions.

We then consider the case that $d \ge 1$. If $\Delta r\left(d-1, d, x^s\left(d\right), c\right) > \gamma$, we then have $\Delta r\left(d_i, d, x^s\left(d\right), c\right) > \gamma$ for all $d_i < d-1$. Suppose user $i$ deletes $\Delta d_i \ge 1$ subscriptions, the loss on its content utility is $\sum_{l=1}^{\Delta d_i} \Delta r\left(d-l, d, x^s\left(d\right), c\right)$, which is larger than $\Delta d_i \gamma$. Therefore, no user will have the incentive to delete existing subscriptions.

Summing up the above discussions, Statement (3) follows. ■

*7) Proof of Theorem 2*

(1) To prove this, we only need to show that $\underline{\gamma}\left(d\right) < \overline{\gamma}\left(d\right)$ always holds for all $d \in \left\{0, \ldots, n-1\right\}$. That is, $\Delta r\left(d, d, x^s\left(d\right), c\right) < \Delta r\left(d-1, d, x^s\left(d\right), c\right)$. This is always true according to Proposition 2 and hence, Statement (1) follows.

(2) To prove $\overline{\gamma}\left(d\right) < \underline{\gamma}\left(d-1\right)$, it is equivalent to show that $\Delta r\left(d-1, d-1, x^s\left(d-1\right), c\right) > \Delta r\left(d-1, d, x^s\left(d\right), c\right)$. This can be proved using a similar approach as Lemma 4. The detailed computation is omitted here due to the space limitation. ■

## Appendix C

*1) Proof of Proposition 3*

(1) Suppose the network is complete with $\overline{g}_{ij}^* = 1,\ \forall i, j$, and there are two users $i$ and $i'$ with $x_i^* > x_{i'}^*$. Denote $y_i^* = \left(x_i^*\right)^\rho + \sum_{j \neq i}\left(x_j^*\right)^\rho$ and $y_{i'}^* = \left(x_{i'}^*\right)^\rho + \sum_{j \neq i'}\left(x_j^*\right)^\rho$ as the amount of contents that user $i$ and $i'$ acquire from others respectively, we have that

$$v'\left(X_i^*\right)\left(\frac{X_i^*}{x_i^*}\right)^{1-\rho} = v'\left(X_{i'}^*\right)\left(\frac{X_{i'}^*}{x_{i'}^*}\right)^{1-\rho} = c\,. \tag{34}$$

However, since $X_i^* = X_{i'}^*$, it can be concluded that $x_i^* = x_{i'}^*$, which contradicts the assumption. Hence this statement follows.

(2) Suppose that there is a $j > n_h$ such that $g_{ji}^* = 0,\ \forall i \leq n_h$. This implies that user $j$ does not establish links with others. If, on the contrary, user $j$ establishes links, then these links are directed to users $j' > n_h$, but then user $j$ can strictly increase its utility by switching a link from $j'$ to some $i \leq n_h$.

Since $g_{ji}^* = 0,\ \forall i \leq n_h$, it can also be claimed that user $j$ does not receive any subscription, i.e. $g_{ij}^* = 0,\ \forall i \neq j$. Suppose user $j$ receives a link from user $j'$, then it must be the case that user $n_h$ is also the neighbor of user $j'$. Otherwise $j'$ can strictly increase his utility by switching the link from $j$ to $n_h$. This implies that every user who establishes a link with user $j$ is also a neighbor of user $n_h$. Regarding the fact that user $j$ only receives subscriptions and does not subscribe to any other, he acquires from his neighbors at most as much contents as user $n_h$ does. Since a user's marginal utility of information production monotonically decreases against the amount of contents he acquires from his neighbors, we can conclude that $x_j^* \geq x_{n_h}^*$, which contradicts the fact that $x_j^* < x_{n_h}^*$. Hence, this statement follows.

(3) When $n_h = 1$, user 1 connects with all the other users in the network according to Statement (2) and thus acquires all contents in the network. Consider another user $j > n_h$, the amount of contents he acquires is always smaller than or equal to the amount of contents that user 1 acquires. Therefore, we have $x_j^* \geq x_1^*$, which contradicts the presumption that $x_j^* < x_1^*$. Hence this statement follows.

(4) If there is a user $j > n_h$, such that $g_{ij}^* = 1$. There is also a user $i' \leq n_h$, such that $\overline{g}_{ii'}^* = 0$. Hence, user $i$ can always monotonically increase his utility by switching the link from $j$ to $i'$. Hence, this statement follows.

(5) Suppose a user $i \leq n_h$ is connected with all $i' \leq n_h$, there is a user $j > n_h$ such that $\overline{g}_{ij}^* = 0$. Clearly, $g_{jj'}^* = 0$ for all $j' > n_h$, otherwise, $j$ strictly increases his utility by switching the link from $j'$ to $i$. Similar to Proposition 4(1), if $j$ receives a link from some $j' > n_h$, $i$ must also be a neighbor of $j'$. Therefore, every friend of user $j$ is also a friend of user $i$ and thus user $i$ acquires more contents than user $j$. We have that $x_j^* \geq x_i^*$, which contradicts the presumption that $x_j^* < x_i^*$. Hence, this statement follows. ■

*2) Proof of Corollary 1*

(1) If $\tilde{x} = \overline{x}$, then each user $i \leq n_h$ should be isolated with no friends according to (7). This contradicts the conclusion of Proposition 3 (2) that there is at least some $i \leq n_h$ that is connected with some $j > n_h$. Hence, this statement follows.

(2) Suppose there exists a user $i \le n_h$ and a user $j > n_h$ such that $\left|\left(X_i^*\right)^\rho - \left(X_j^*\right)^\rho\right| \ge \left(\tilde{x}\left(\boldsymbol{x},\boldsymbol{g}\right)\right)^\rho$ in an asymmetric equilibrium. As Proposition 3 (5) shows, each user $i \le n_h$ does not connect to at least one user $i' \le n_h$ at equilibrium. His content utility can be represented as $r_i\left(\tilde{x}, x_{-i}^*, \boldsymbol{g}^*\right) = v\left(X_i^*\right) - c\tilde{x}$. Now consider the situation when user $i$ subscribes to user $i'$ and denote his optimal amount of content production as $x_i$. His content utility is then $r_i\left(x_i, x_{-i}^*, \boldsymbol{g}'\right) = v\left(\left(X_i^*\right)^\rho + \left(x_i\right)^\rho\right) - cx_i$, where $\boldsymbol{g}'$ is the same to $\boldsymbol{g}^*$ except for the entry $g'_{ii'} = 1$. Since user $i$ has no incentive to subscribe to $i'$, we have that $r_i\left(x_i, x_{-i}^*, \boldsymbol{g}'\right) - r_i\left(\tilde{x}, x_{-i}^*, \boldsymbol{g}^*\right) < \gamma$.

Proposition 3 (1) shows that each user $j > n_h$ subscribes to a user $i'' \le n_h$. His content utility is $r_j\left(x_j^*, x_{-j}^*, \boldsymbol{g}^*\right) = v\left(X_j^*\right) - cx_j^*$. Now consider the situation when $j$ deletes the subscription to $i''$ and let $x_j$ denote his optimal content production. His content utility now becomes $r_j\left(x_j, x_{-j}^*, \boldsymbol{g}''\right) = v\left(X_j^*\right) - cx_j$, where $\boldsymbol{g}''$ is the same to $\boldsymbol{g}^*$ except for the entry $g''_{ji''} = 0$. Since user $j$ has no incentive to delete the subscription in equilibrium, we have that $r_j\left(x_j^*, x_{-j}^*, \boldsymbol{g}^*\right) - r_j\left(x_j, x_{-j}^*, \boldsymbol{g}''\right) > \gamma$.

Expanding $r_i\left(x_i, x_{-i}^*, \boldsymbol{g}'\right) - r_i\left(\tilde{x}, x_{-i}^*, \boldsymbol{g}^*\right)$, we have that

$$
\begin{aligned}
& r_i\left(x_i, x_{-i}^*, \boldsymbol{g}'\right) - r_i\left(\tilde{x}, x_{-i}^*, \boldsymbol{g}^*\right) \\
& = r_i\left(x_i, x_{-i}^*, \boldsymbol{g}'\right) - \left(v\left(\left(\left(X_i^*\right)^\rho + \tilde{x}^\rho\right)^{\frac{1}{\rho}}\right) - c\tilde{x}\right) + \left(v\left(\left(\left(X_i^*\right)^\rho + \tilde{x}^\rho\right)^{\frac{1}{\rho}}\right) - c\tilde{x}\right) - r_i\left(\tilde{x}, x_{-i}^*, \boldsymbol{g}^*\right) \\
& > \left(v\left(\left(\left(X_i^*\right)^\rho + \tilde{x}^\rho\right)^{\frac{1}{\rho}}\right) - c\tilde{x}\right) - r_i\left(\tilde{x}, x_{-i}^*, \boldsymbol{g}^*\right) \\
& \geq \left(v\left(X_j^*\right) - cx_j^*\right) - \left(v\left(\left(\left(X_j^*\right)^\rho - \tilde{x}^\rho\right)^{\frac{1}{\rho}}\right) - cx_j^*\right) \\
& > \left(v\left(X_j^*\right) - cx_j^*\right) - \left(v\left(\left(\left(X_j^*\right)^\rho - \tilde{x}^\rho - \left(x_j^*\right)^\rho + \left(x_j\right)^\rho\right)^{\frac{1}{\rho}}\right) - cx_j\right) \\
& = r_j\left(x_j^*, x_{-j}^*, \boldsymbol{g}^*\right) - r_j\left(x_j, x_{-j}^*, \boldsymbol{g}''\right) > \gamma
\end{aligned}
\tag{35}
$$

This contradicts our conclusion that $r_i\left(x_i, x_{-i}^*, \boldsymbol{g}'\right) - r_i\left(\tilde{x}, x_{-i}^*, \boldsymbol{g}^*\right) < \gamma$ . Hence, this proposition follows. ■

*3) Proof of Corollary 2*

(1) Suppose a star topology with the center being user $n$ and the spokes being agents $1 \sim n-1$ being an equilibrium. Due to the symmetry of the topology, we have that $x_1^* = \ldots = x_{n-1}^*$, which is denoted as $\tilde{x}\left(\boldsymbol{x}^*, \boldsymbol{g}^*\right)$. The equilibrium production $x_n^*$ of user $n$ is strictly small than $\tilde{x}\left(\boldsymbol{x}^*, \boldsymbol{g}^*\right)$, which is denoted as $\underset{\sim}{x}\left(\boldsymbol{x}^*, \boldsymbol{g}^*\right)$ . Hence, the perceived amount of contents for user $n$ is $X_n^* = \left(\left(\underset{\sim}{x}\left(\boldsymbol{x}^*, \boldsymbol{g}^*\right)\right)^\rho + \left(n-1\right)\left(\tilde{x}\left(\boldsymbol{x}^*, \boldsymbol{g}^*\right)\right)^\rho\right)^{\frac{1}{\rho}}$, and the perceived amount of contents for user $j < n$ is $X_j^* = \left(\left(\underset{\sim}{x}\left(\boldsymbol{x}^*, \boldsymbol{g}^*\right)\right)^\rho + \left(\tilde{x}\left(\boldsymbol{x}^*, \boldsymbol{g}^*\right)\right)^\rho\right)^{\frac{1}{\rho}}$ . We have that $\left|\left(X_n^*\right)^\rho - \left(X_j^*\right)^\rho\right| \geq \left(\tilde{x}\left(\boldsymbol{x}^*, \boldsymbol{g}^*\right)\right)^\rho$ . According to Corollary 1 (2), the star topology cannot be an equilibrium, since the content distribution is highly unbalanced with the central user accesses far more contents than spoke users. Hence, this statement follows.

(2) Without loss of generality, we assume $n$ to be odd. For notational convenience, the user in the center of the line, i.e. the $\frac{n+1}{2}$-*th* user from any end, is called user $1$. We also index the pair of users whose distance to user 1 is $d$ using $2d$ and $2d+1$ for $1 \le d \le \frac{N-1}{2}$, with the left user being $2d$ and the right user being $2d+1$. Due to the symmetry, we only have to analyze users at the left half of the line.

When $n=3$, the line also forms a star topology and hence cannot be sustained according to Statement (1).

When $n \ge 5$, it is obvious that the perceived amount of information that user $n-1$ acquires is strictly lower than user $n-3$. Hence, $x_{n-1}^*$ is strictly higher than $x_{n-3}^*$. Since user $n-1$ is only connected with user $n-3$ which is not a user with the highest information production, we should have $x_{n-1}^*$ being the highest amount of information production according to Proposition 3 (2).

If $n=5$, we should have $x_1^* = x_{n-1}^*$ since user 1 is connected to neither user $n-1$ nor user $n$. Hence, the perceived amount of contents of user 1 is strictly higher than user $n-1$, which contradicts the fact that $x_1^* = x_{n-1}^*$. Therefore, the line cannot be sustained in equilibrium.

If $n \ge 7$, it should be the case that (1) $x_{n-5}^* = x_{n-1}^*$; or (2) $x_{n-5}^* < x_{n-1}^*$ and $x_{\max\{n-7,1\}}^* = x_{n-1}^*$. The first case does not hold due to the same reason as in the scenario when $n=5$. In the second case, we have the perceived amount of contents that user $n-5$ acquires equals to that of user $n-3$ and hence $x_{n-5}^* = x_{n-3}^* < x_{n-1}^*$. Since $\overline{g}_{n-3,n-5}^* = 1$, we assume $g_{n-3,n-5}^* = 1$ without loss of generality. As a result, user $n-3$ can strictly increases his utility if he switches the subscription from user $n-5$ to user $\max\{n-7,1\}$, which contradicts the equilibrium property. Therefore, the line cannot be sustained in equilibrium and Statement (2) follows. ■

*4) Proof of Theorem 3*

For illustration purposes, let $n_l = n - n_h$ denote the population of low producers. Since each low producer subscribes to at least one high producer, we should have the following inequality as

$$v\left(\tilde{x}\left(\boldsymbol{x}^*, \boldsymbol{g}^*\right)\right) > \gamma, \tag{36}$$

in order to sustain a low producer's incentive to create a link. Hence, $\tilde{x}$ is lower bounded as follows:

$$\tilde{x}\left(\boldsymbol{x}^*, \boldsymbol{g}^*\right) > \frac{\gamma}{\alpha}. \tag{37}$$

For any value of $\tilde{x}\left(\boldsymbol{x}^*, \boldsymbol{g}^*\right)$ which satisfies (37), there exists an integer $L_h$ such that

$$v\left(\left(L_h + 1\right)^{\frac{1}{\rho}} \tilde{x}\left(\boldsymbol{x}^*, \boldsymbol{g}^*\right)\right) - v\left(L_h^{\frac{1}{\rho}} \tilde{x}\left(\boldsymbol{x}^*, \boldsymbol{g}^*\right)\right) < \gamma. \tag{38}$$

Consequently, a low producer has no incentive to subscribe to more than $L_h$ high producers, and there are only two situations that will possibly emerge when $n \to \infty$:

(1) The number of high producers remains finite, i.e. $n_h < \infty$ when $n \to \infty$.

(2) $n_h$ goes to infinity when $n \to \infty$ and there is no low producer who subscribes to other low producers. Alternatively speaking, each low producer only subscribes to high producers.

To analyze the first scenario, we further classify low producers described in Proposition 4 into the following classes:

(a) A low producer who only subscribes to high producers and is not subscribed by other low producers;

(b) A low producer who only subscribes with high producers and is also subscribed by other low producers;

(c) A low producer who subscribes to all high producers and some low producers.

For notational convenience, the population sizes of the above three classes are denoted as $n_{la}$, $n_{lb}$, and $n_{lc}$, respectively.

It is obvious that production level of a low producer of type (a) is bounded away from 0 since the total amount of contents he acquires will not exceed a finite amount $n_h \bar{x}$. If $n_{la}$ goes to infinity, there is at least one high producer being subscribed by an infinite amount of users of type (a). Hence, the amount of contents produced by this high producer should go arbitrarily close to 0, which leads to a contradiction to (37). As a result, it can be concluded that $n_{la}$ is finite when $n \to \infty$.

Now consider a low producer $j$ who belongs to type (b). Let $x_j^*$ denote his production level and we should also have $x_j^*$ satisfying (37). Otherwise, no user has the incentive to subscribe to him. Using the same argument as that for type (a), it can be concluded that $n_{lb}$ is also finite when $n \to \infty$.

To analyze low producers of type (c), we further classify them into the following two sub-classes and denote their populations as $n_{lc1}$ and $n_{lc2}$, respectively.

(c1) a low producer of type (c) who produces an amount of contents which satisfy (37);

(c2) a low producer of type (c) who produces an amount of contents which does not satisfy (37).

It is obvious that $n_{lc1}$ is also upper-bounded when $n \to \infty$. Since $n_h$, $n_{la}$, $n_{lb}$, and $n_{lc1}$ are all finite when $n \to \infty$, $n_{lc2}$ should go to infinity. Meanwhile, we have that any two users of type (c2) do not mutually subscribe to each other since the subscription cost exceeds the benefit they can obtain from mutual content sharing. Therefore, the maximum number of friends that a user of type (c2) has is $n_h + n_{la} + n_{lb} + n_{lc1}$, which is finite when $n \to \infty$. As a result, the production level of a user of type (c2) is lower-bounded away from 0. Because each user of type (c2) subscribes to all high producers and $n_{c2}$ goes to infinity, the amount of contents produced by each high producer should still go arbitrarily close to 0, which leads to a contradiction to (37). Hence $n_{lc2}$ should also be finite and the first scenario, i.e. $n_h < \infty$ when $n \to \infty$, cannot emerge in any asymmetric equilibrium when $n \to \infty$.

Next we analyze the second scenario as $n_h$ goes to infinity together with $n$ and there are no two low producers who mutually subscribe to each other, i.e $\overline{g}^*_{jj'} = 0$ for any $j, j' > n_h$. Consider two low producers $j$ and $j'$. Let $d_j$ and $d_{j'}$ denote their degrees in equilibrium, respectively. Without loss of generality, we assume that $d_j > d_{j'}$, or alternatively, $d_j \geq d_{j'} + 1$. Since user $j$ has no incentive to delete his existing subscriptions, we have that

$$r_j\left(x_j^*, x_{-j}^*, \boldsymbol{g}^*\right) - r_{j'}\left(x_{j'}^*, x_{-j'}^*, \boldsymbol{g}^*\right) > \left(d_j - d_{j'}\right)\gamma . \tag{39}$$

Similarly, since user $j'$ has no incentive to add new subscriptions, we have that

$$r_j\left(x_j^*, x_{-j}^*, \boldsymbol{g}^*\right) - r_{j'}\left(x_{j'}^*, x_{-j'}^*, \boldsymbol{g}^*\right) < \left(d_j - d_{j'}\right)\gamma . \tag{40}$$

Hence, there is a contradiction and each low producer should subscribe to the same number of high producers, denoted as $\underset{\sim}{k}\left(\boldsymbol{x}^*, \boldsymbol{g}^*\right)$, and has the same production level, denoted as $\underset{\sim}{x}\left(\boldsymbol{x}^*, \boldsymbol{g}^*\right)$. Similarly, since high producers have the same production level, they should also subscribe to the same number of high producers, denoted as $\tilde{k}\left(\boldsymbol{x}^*, \boldsymbol{g}^*\right)$. Hence, Theorem 3 follows. ■

*5) Proof of Theorem 4*

It can be learned from (38) that a low content producer cannot connect to more than $L_h$ high content producers, which upper-bounds the amount of contents he receives from others at $L_h \overline{x}$. Hence, the production level $\underset{\sim}{x}\left(\boldsymbol{x}^*, \boldsymbol{g}^*\right)$ of a low content producer is bounded away from 0.

If $\lim_{n \to \infty} \frac{n_h}{n_l} = 0$, we can always find a sufficiently large $n$ such that each high content producer receives a sufficiently large number of subscriptions and thus a sufficiently large amount of contents from low content producers, which leads to a production level $\tilde{x}\left(\boldsymbol{x}^*, \boldsymbol{g}^*\right) < \underset{\sim}{x}\left(\boldsymbol{x}^*, \boldsymbol{g}^*\right)$. This contradicts the fact that $\tilde{x}\left(\boldsymbol{x}^*, \boldsymbol{g}^*\right) > \underset{\sim}{x}\left(\boldsymbol{x}^*, \boldsymbol{g}^*\right)$ and hence Theorem 4 follows. ■

*6) Proof of Corollary 3*

(1) Suppose there are an influencer $i$ who subscribes to $t_h$ influencers and a subscriber $j$ who subscribes to $t_l$ influencers. We also assume that the utility received by $j$ is higher than that received by $i$. As a result, if $i$ reduces his production level to $\underset{\sim}{x}\left(\boldsymbol{x}^*, \boldsymbol{g}^*\right)$ and adds the number of his subscriptions to $t_l$, he can always receive a utility that is higher than that received by $j$, since he can acquire the additional contents produced by the subscribers who subscribe to him, which is in turn higher than the utility he receives when he produces an amount $\tilde{x}\left(\boldsymbol{x}^*, \boldsymbol{g}^*\right)$ and has $t_h$ subscriptions. This leads to a contradiction to the equilibrium condition and hence Statement (1) follows.

(2) It is obvious that the perceived amount of contents of an influencer is always lower than that of a subscriber. Suppose an influencer $i$ has $t_h$ links with $t_h \geq t_l$, compared to an arbitrary subscriber $j$, $i$ has a lower perceived amount of contents, but is subject to a higher production and subscription cost. As a result, $i$'s utility is always smaller than $j$, which leads to a contradiction to Corollary 3. Hence, Statement (2) follows. ■

## Appendix D

*1) Proof of Theorem 5*

Consider two users $i$ and $j$ who are friends with each other. Meanwhile, we assume that $X_i > X_j$ under a strategy profile that has achieved the social optimum. Regarding both $X_i$ and $X_j$ as functions of $x_i$, we should have $c = \frac{\partial v\left(X_i\right)}{\partial x_i} < \frac{\partial v\left(X_j\right)}{\partial x_i}$. Therefore, there is a value $x_i' > x_i$, such that

$$\begin{aligned} & v\left(\left(\left(X_i\right)^\rho - x_i^\rho + \left(x_i'\right)^\rho\right)^{\frac{1}{\rho}}\right) + v\left(\left(\left(X_j\right)^\rho - x_i^\rho + \left(x_i'\right)^\rho\right)^{\frac{1}{\rho}}\right) - cx_i' \\ & > v\left(X_i\right) + v\left(X_j\right) - cx_i \end{aligned} . \tag{41}$$

Since the utility of any other user will not be influenced by such adjustment, the social welfare monotonically increases. Hence, we should have $X_i = X_j$ for any two users $i$ and $j$ within the same component. It is straightforward that users within the same component have the same degree and the same production level.

Now assuming that there are $m > 1$ components under a strategy profile $\boldsymbol{s} = \left(\boldsymbol{x}, \boldsymbol{g}\right)$, which are indexed from 1 to $m$. Suppose users from different components have different utilities and assume that users from component 1 have the highest individual utility without loss of generality. Then, it is straightforward that users from components other than 1 can join component 1 with the social welfare monotonically increasing. Hence, it should be concluded that users from different components should have the same individual utility. Therefore, the resulting social optimum can always be achieved by a symmetric profile where only one component exists in the network. ■

*2) Proof of Proposition 5*

Taking the partial derivative in $d$ over (21), we have

$$\begin{aligned} & \frac{\partial \Delta q\left(x^{\#}\left(d\right), d\right)}{\partial d} = \frac{\partial\left[q\left(x^{\#}\left(d+1\right), d+1\right) - q\left(x^{\#}\left(d\right), d\right)\right]}{\partial d} \\ & = \frac{\partial\left[v\left(\left(2+d\right)x^{\#}\left(d+1\right)\right)\right]}{\partial d} - \frac{\partial\left[v\left(\left(1+d\right)x^{\#}\left(d\right)\right)\right]}{\partial d} - c\left(\frac{\partial x^{\#}\left(d+1\right)}{\partial d} - \frac{\partial x^{\#}\left(d\right)}{\partial d}\right) . \\ & = \frac{1}{2+d} x^{\#}\left(d+1\right) - \frac{1}{1+d} x^{\#}\left(d\right) \end{aligned} \tag{42}$$

where the second equality applies (19). Similar to Lemma 4, we have $x^{\#}\left(d\right) > x^{\#}\left(d+1\right)$ and hence $\frac{\partial \Delta q\left(x^{\#}\left(d\right), d\right)}{\partial d} < 0$ . Therefore, Proposition 5 follows. ■